\documentclass[final,9p,times]{elsarticle}
\usepackage[a4paper,margin=1in]{geometry}

\usepackage[labelfont=bf, labelsep=period]{caption} 
\usepackage{setspace}
\usepackage{lineno}
\usepackage{graphicx}
\usepackage{amssymb}
\usepackage{newtxtext}
\usepackage{newtxmath}
\usepackage{natbib}
\usepackage{xcolor}
\usepackage{hyperref}
\usepackage{cleveref} 
\usepackage{float}

\usepackage{amsmath}
\usepackage{booktabs}
\usepackage{tabularx}
\usepackage{array} 
\usepackage{caption}  
\usepackage{algorithm} 
\usepackage{algpseudocode}
\hypersetup{
    colorlinks=true,
    citecolor=blue,   
    linkcolor=blue,    
    filecolor=blue,   
    urlcolor=blue      
}

\usepackage[normalem]{ulem}

\crefname{equation}{equation}{Eqs.}
\Crefname{equation}{equation}{Eqs.}
\crefname{figure}{Fig.}{Figs.}
\Crefname{Figure}{Fig.}{Figs.}
\crefname{algorithm}{Algorithm}{Algorithms}
\Crefname{algorithm}{Algorithm}{Algorithms}
\crefname{section}{Section}{Sections}
\Crefname{Section}{Section}{Sections}
\crefname{table}{Table}{Tables}
\Crefname{Table}{Table}{Tables} 

\begin{document}

\begin{frontmatter}

\title{Vortex shedding suppression in elliptical cylinder via reinforcement learning}

\author{Wang Jia, Hang Xu\textsuperscript{*}}

\affiliation{organization={State Key Laboratory of Ocean Engineering, School of Ocean and Civil Engineering, Shanghai Jiao Tong University},
            city={Shanghai},
            postcode={200240}, 
            country={China}}

\begin{abstract}
\noindent
Flow control of bluff bodies plays a critical role in engineering applications. 
In this study, deep reinforcement learning (DRL) is employed to develop flow control strategies for the flow past an elliptical cylinder confined between two walls.
The primary objective is to investigate the feasibility of achieving multi-objective flow control for an elliptical cylinder with varying aspect ratios ($Ar$), while maintaining low control energy input.
DRL training results demonstrate that for an elliptical cylinder with larger $Ar$, the control strategy effectively reduces drag, minimizes lift fluctuations, and completely suppresses vortex shedding, all while maintaining low external energy consumption. 
Conversely, decreasing the $Ar$ compromises the effectiveness of multi-objective control, even when greater energy input is applied.
Through detailed physical analysis, the coupling effect between the blockage ratio ($\beta$) and $Ar$ is identified as a limiting factor for vortex shedding suppression and wake stabilization.
At lower values of $\beta$, the control strategy successfully achieves multi-objective optimization for elliptical cylinders across the entire range of $Ar$.
Although balancing energy efficiency and control performance remains challenging for highly slender cylinders, the proposed DRL strategy still achieves effective vortex shedding suppression.
This work highlights the potential of DRL-based control strategies to effectively stabilize wake flows around slender bluff bodies, with an explicit emphasis on maintaining energy efficiency.

\end{abstract}


\end{frontmatter}

\section{Introduction}   

Since the time of Strouhal in 1878, the vortex shedding phenomenon in bluff body flows has been a subject of extensive research and interest \cite{Strykowski1990,Graham2024}.
Vortex shedding causes vibrations, increasing drag, fatigue, collision risks, and noise, compromising performance and safety \cite{Cicolin2024,Matheswaran2024}.
To mitigate these effects, researchers have developed control technology that are broadly categorized as active flow control (AFC) or passive flow control \cite{SCOTTCOLLIS2004237,fluids9120290,li2025formation}. 
Among these, AFC offers greater flexibility and tunability, enabling dynamic adjustments to real-time flow conditions for more precise control \cite{LI2022106034,Mali2025}. 
In 2019, Rabault et al. \cite{rabault2019artificial} pioneered the integration of reinforcement learning with synthetic jet actuation for cylinder flow control, catalyzing subsequent advancements in: sensor placement optimization \cite{parisRobust}, enhanced adaptability in various bluff-body flow environments \cite{jia2024robust,JIA2025111125}, extension to higher Reynolds numbers \cite{Varela2022,jia2024jetsactuator,JIA2025109203} and three-dimensional flows \cite{Amico2022,Font2025turbulent}, and application in physical experiments \cite{Dixiapnas,Zong2025}.

Despite extensive research in flow control, studies on slender structures with an extremely small aspect ratio ($Ar$) remain limited.
The wake dynamics of elliptical cylinders are inherently more complex than those of circular cylinders, exhibiting greater instability, stronger vortex interactions, and distinct flow transitions \cite{Thompson2014,GURSUL201417,Rockwell}. 
These complexities pose additional challenges for flow control, requiring strategies that effectively adapt to intricate wake dynamics.
In this study, we address two pivotal challenges: (i) achieving complete vortex suppression through flow control on elliptical cylinders, and (ii) analyzing the sensitivity of control effectiveness to $Ar$ and blockage ratio ($\beta$) based on physical insights. Particular focus is placed on investigating the potential of model-free control strategies to completely suppress vortex shedding in bluff bodies, thereby laying the groundwork for more advanced flow control applications.

\subsection{Advancing fluid dynamics through machine learning}

Traditionally, fluid mechanics has relied on empirical laws or semi-empirical models to describe and predict fluid behavior \cite{Ali2021,LI2024109110}. 
However, with the rapid development of big data and machine learning, a paradigm shift has emerged, wherein models are constructed by analyzing large datasets \cite{feng2025typhoon,zhao2025comparative}. 
Unlike traditional approaches that depend on predefined empirical laws, data-driven modeling learns directly from data, enabling the accurate representation and prediction of fluid behavior \cite{jiang2025koopman,jiang2024balanced}. 
This data-driven approach not only enhances prediction and control efficiency but also overcomes the inherent limitations of conventional methods, offering new possibilities for advancing fluid dynamics \cite{annurevMehdi,Brunton2024}.

The application of machine learning into fluid mechanics has revolutionized various key areas, including modeling \cite{jiang2025koopman,subgridscale}, optimization \cite{BHOLA2023112018}, prediction \cite{LICheng2025,LUO2024118493,feng2024multistation}, and control \cite{rabault2019artificial}. 
This integration has significantly bolstered the efficiency and effectiveness of fluid dynamics problem-solving \cite{GARNIER,Viquerat,rabault2020deep}.
Machine learning aids in identifying low-dimensional manifolds and flow states, enabling a deeper understanding of  flow phenomena \cite{HIGASHIDA2024243}.
In reduced-order modeling and shape optimization, it efficiently accelerates optimal design \cite{LU20211351}.
Moreover, machine learning enables rapid and accurate flow condition predictions \cite{VINOKIC2025100297}.
In flow control, it dynamically adjusts flow parameters in real time to optimize control performance \cite{jia2024robust}.
Collectively, these applications underscore the transformative impact of machine learning in fluid mechanics.

\subsection{Applications of deep reinforcement learning}

Reinforcement learning (RL) is an advanced machine learning paradigm that learns optimal strategies through continuous interaction with the environment, eliminating the need for large amounts of labeled data \cite{Kaelbling1996}. 
Its key strengths—autonomous learning, adaptability, and long-term strategy optimization—enable applications in diverse fields such as robotics, autonomous driving, and medical diagnosis \cite{granter2017alphago,HEUILLET2021106685,Ma2024,Iwami2022}.
Deep reinforcement learning (DRL) enhances RL by integrating deep neural networks (DNNs), which effectively approximate nonlinear functions in high-dimensional spaces \cite{LIU201711,Mnih2015}. 
DRL leverages the powerful feature extraction and high-dimensional processing capabilities of DNNs to overcome key limitations of traditional RL. This enables more efficient learning in complex environments \cite{9904958}.

In fluid dynamics, governing equations often exhibit significant nonlinearity and high dimensionality, particularly in turbulent and chaotic flows, which limits the effectiveness of traditional linearized control methods \cite{Ali2021,Font2025}. 
In contrast, black-box RL methods excel in controlling such complex systems without requiring explicit physical models, instead learning optimal strategies through dynamic interaction with the environment \cite{Kaelbling1996,franccois}.
RL advances fluid mechanics by improving turbulence prediction \cite{largeeddy}, refining boundary layer modeling \cite{subgridscale,wallmodels}, and accelerating CFD solver convergence \cite{Niemeyer2014,huergo2024}. Additionally, it enhances aerodynamic design by optimizing airfoils and vehicle shapes for reduced drag \cite{KIM2022111263,LI2022100849}. 
In flow control, RL effectively regulates wakes around bluff bodies and reduces friction drag in turbulent channels, contributing to improved efficiency and performance \cite{Sonoda2023,Ciri2021}. 
These applications underscore the critical role that RL plays in control and optimization across fluid mechanics \cite{Viquerat,annurevfluid}.

\subsection{Deep reinforcement learning-based active flow control}

Significant advancements in DRL-based active flow control have been made through the sustained and collaborative efforts of the research community. The development and widespread adoption of open-source DRL and CFD coupling codes \cite{rabault2019artificial, liReinforcementlearning, wangDRLinFluids}, along with the creation of robust open-source reinforcement learning frameworks \cite{wangDRLinFluids,KURZ}, have established a solid and versatile foundation for continued progress in this field.
The pioneering work by Rabault et al. marked the introduction of Proximal Policy Optimization (PPO) to active flow control, demonstrating its capability to learn optimal strategies for cylinder flow at $Re=100$ by adjusting jet mass flow rates \cite{rabault2019artificial}. 
This foundational work was further extended by Rabault et al. \cite{rabaultAccelerating}, who improved parallelization techniques, and Jia et al. \cite{jia2024optimal}, who achieved a remarkable 47-fold speedup in computational efficiency.

In terms of control strategies, Jia et al.  \cite{jia2024robust} applied the Soft Actor-Critic (SAC) algorithm for multi-objective control of square cylinders, while Xia et al. \cite{Xia2024} proposed a dynamic RL scheme designed to reduce drag under partial measurements. 
Wang et al. \cite{Dynamic2024} developed a feature-based DRL algorithm that effectively handles control with sparse sensor data. 
Further advancements in control effectiveness have been made through the optimization of probe placement and quantity \cite{parisRobust, liReinforcementlearning, Dynamic2024}, as well as sensitivity analyses of jet positioning \cite{jia2024jetsactuator, yan2023stabilizing}. Additionally, the transition from synthetic jets to alternative actuation methods \cite{Feng2023} and the extension of DRL-based control to slender structures \cite{Wang2024} have demonstrated the growing versatility and applicability of these techniques.
\Cref{tab:table1} summarizes the effectiveness of DRL-based AFC approaches across various scenarios, including the use of twin delayed deep deterministic policy gradient (TD3) and truncated quantile critics (TQC).

\begin{table*}[hbt!]
\centering
\caption{Comparative review of recent applications of deep reinforcement learning in active flow control of bluff body wakes. 
The table summarizes key parameters including the Reynolds number, bluff body geometry, actuation mechanism, achieved control performance, DRL algorithm employed, numerical solver used, and relevant references.
}
\resizebox{\textwidth}{!}{ 
\begin{tabular}{lccccccc}
\toprule
    $Re$ & Bluff Body & Method & Control Performance & Control Algorithm & Solver & Reference \\[3pt]
\hline    
    100 & Cylinder & Synthetic Jets & 8\% drag reduction & PPO & \texttt{FEniCS} & Rabault et al. \cite{rabault2019artificial} \\
    200 & Cylinder & Oscillating cylinder & 8\% drag reduction & PPO & \texttt{Nek5000} &  Jiang et al. \cite{jiang2023reinforcement} \\      
    100 & Square & Synthetic Jets & 17\% drag reduction & TQC & \texttt{FEniCS} & Xia et al. \cite{Xia2024} \\    
    400 & Square & Synthetic Jets & 47\% drag reduction & SAC & \texttt{OpenFOAM} & Jia et al. \cite{jia2024robust} \\
    100 & Pinball & Cylinder Rotation & Force extremum and tracking & TD3 & \texttt{BDIM} & Feng et al. \cite{Feng2023} \\   
    100 & Ellipse & Synthetic Jets & 44\% drag reduction & PPO & \texttt{OpenFOAM} & Jia et al. \cite{Wang2024} \\   
\bottomrule
\end{tabular}
}
\label{tab:table1}
\end{table*}

To address more complex flow scenarios, recent studies have extended DRL-based control strategies to higher Reynolds numbers and three-dimensional flows.
Varela et al. \cite{Varela2022} reported that increasing the Reynolds number significantly alters DRL-derived control strategies, while Guastoni et al. \cite{Guastoni2023} demonstrated the effectiveness of DRL in discovering drag reduction strategies for three-dimensional channel flows.
Furthermore, Font et al. \cite{Font2025} compared classical control methods with DRL-based approaches for minimizing turbulent separation, highlighting the computational efficiency and superior performance of DRL.
In parallel, the practical application of DRL in industrial settings has gained traction \cite{Amico2022,Zong2025,dong2024interactive}.
Amico et al. \cite{Amico2022} made significant progress by implementing DRL-based AFC in experiments to manage three-dimensional bluff body wakes at high Reynolds numbers.
Similarly, Dong et al. \cite{dong2023surrogate} addressed challenges in experimental DRL-based AFC using surrogate models.

\subsection{Objectives of this work}

The wake of an elliptic cylinder exhibits complex transitions, including dual-layer shedding, secondary vortex streets, and intensified interactions, resulting in greater instability compared to a circular cylinder \cite{Amini2024,Lo2024}. These challenges are exacerbated for highly slender bodies, where flow sensitivity and instability complicate effective control.
Wang et al. \cite{Wang2022Active} applied DRL to control the flow around elliptical cylinders with an $Ar$ greater than 1. 
Their study found that achieving a higher drag reduction invariably resulted in flow instability.
In contrast, Jia et al. \cite{Wang2024} conducted a comprehensive study on elliptic cylinders with $Ar$ ranging from 0 to 2, achieving vortex shedding suppression. Their findings indicate that for $Ar$ between 0.75 and 2, DRL-based control can completely suppress vortex shedding, consistent with the results of Wang et al. \cite{Wang2022Active}. 
However, for the flow around highly slender blunt bodies, achieving complete suppression of vortex shedding presents significant challenges, highlighting the complexities inherent in designing effective control strategies.

To address these challenges, our research attempts to solve the problem of stabilizing flow and suppressing vortex shedding in blunt body.
To explore whether it is possible to manipulate the synthetic jet based on the DRL algorithm to achieve efficient, stable and energy-saving flow control effects in the case of blunt-body flow around an elliptical cylinder with an extreme slenderness ratio.
Our focus is on real-time adaptive control techniques for high-dimensional nonlinear flow systems, with an emphasis on improving energy efficiency while maintaining control performance. 
Ultimately, a comprehensive analysis of flow control performance based on physical phenomena is conducted, fostering a deeper understanding and application of flow control mechanisms.

This paper is structured as follows. \Cref{sec:Methodology} outlines our research methods, encompassing both numerical simulations and the DRL framework. \Cref{sec:results} delves into a comprehensive analysis of the results derived from DRL training, with a particular emphasis on elucidating the underlying control mechanisms. Lastly, \Cref{sec:Conclusions} encapsulates the primary findings and their broader ramifications.

\section{Methods}\label{sec:Methodology}
We implement flow control based on DRL for elliptical cylinders with aspect ratios ranging from 0.1 to 1, considering two different $\beta$.
This section outlines the core components of DRL algorithms, focusing on the environment and agent within a flow control framework. \Cref{sec:Simulation} introduces numerical simulation as the environment and details the employed methods. \Cref{sec:ANN} explores the integration of artificial neural networks with RL, emphasizing the PPO algorithm. Finally, \Cref{sec:DRL_AFC} constructs a DRL-based flow control framework and describes its key components.

\begin{figure*}[hbt!]
    \includegraphics{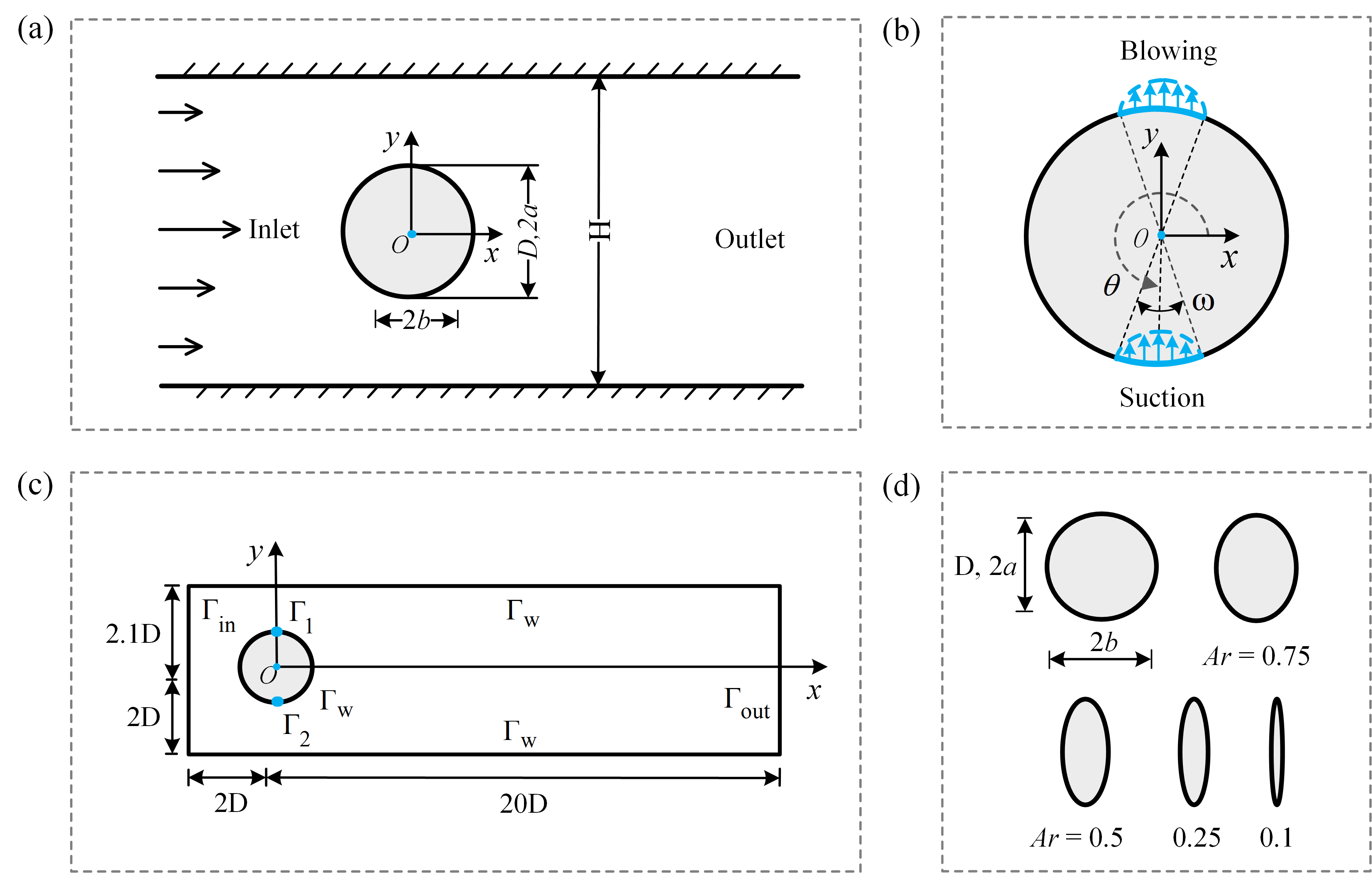}
    \caption{Schematic of the computational setup and the elliptical cylinder configurations used in this study. 
    (a) Computational domain illustrating the two-dimensional flow around an elliptical cylinder in a confined channel. The domain length and height are selected to ensure fully developed inflow and negligible outlet effects. 
    (b) Synthetic jet configuration applied to the cylinder surface. The actuation is driven by a pair of opposing synthetic jets, characterized by the actuation angle $\theta$, that alternate between blowing and suction. 
    (c) Boundary conditions imposed on the computational domain. $\Gamma_{\text{in}}$, $\Gamma_{\text{out}}$, and $\Gamma_{w}$ denote the inlet, outlet, and wall boundaries, respectively. The top and bottom actuation slots located at $\theta = 90°, 270°$ on the cylinder surface are denoted as $\Gamma_1$ and $\Gamma_2$. 
    (d) Elliptical cylinder geometries with different aspect ratios ($Ar = 1.0$, $0.75$, $0.5$, $0.25$, and $0.1$), representing a transition from a circular cylinder to highly slender bodies. The variation in $Ar$ enables assessment of geometric effects on flow control performance.} 
    \label{fig:figure1}
\end{figure*}

\subsection{Simulation environment}\label{sec:Simulation} 

We discuss the flow around a two-dimensional elliptical cylinder in a confined space, as illustrated in \Cref{fig:figure1}(a). 
The physical model is described in the Cartesian coordinate system, with the origin located at the center of the elliptical cylinder.
The $x$-axis aligns with the flow direction, while the $y$-axis represents the direction perpendicular to the flow, which is also the wall-normal direction. 
The synthetic jets are positioned at the 90\textdegree~and 270\textdegree~locations on the elliptical cylinder, as shown in \Cref{fig:figure1}(b).
Length is non-dimensionalized by the cylinder diameter $D$, and velocity is non-dimensionalized by the maximum parabolic inflow velocity $U_{max}$, which will be discussed later.

The computational domain for the numerical simulations extends 20$D$ in the positive $x$-direction and 2$D$ in the negative $x$-direction from the center of the elliptical cylinder. In the $y$-direction, the domain extends 2.1$D$ in the positive direction and 2$D$ in the negative direction. Thus, the computational domain is defined as a rectangular region of 22$D$ × 4.1$D$, as illustrated in \Cref{fig:figure1}(c).
We define the $Ar$ to characterize the shape of the elliptical cylinder, where $Ar$ is given by the ratio of the semi-minor axis $b$ to the semi-major axis $a$, i.e., $Ar = b/a$, as shown in \Cref{fig:figure1}(d).
When the $Ar$ of the elliptical cylinder equals 1, the geometry corresponds to a circular cylinder. When $Ar$ is less than 1, the geometry represents an elliptical cylinder. 
Additionally, the $\beta = D/H$ quantifies the degree of flow confinement.

The Navier-Stokes equations for an incompressible, viscous fluid within a domain $\Omega \subset \mathbb{R}^{nd}$ over a time interval $(0, T)$ form the core of fluid dynamics analysis. 
These equations describe the time evolution of the fluid velocity field $\mathbf{u} = \mathbf{u}(\mathbf{x}, t)$ and the pressure field $p = p(\mathbf{x}, t)$, where $\mathbf{x}$ represents the spatial coordinates and $t$ denotes time.
\begin{subequations}
\begin{equation}
    \frac{\partial \mathbf{u}}{\partial t} + \mathbf{u} \cdot \nabla \mathbf{u} = -\nabla p + \frac{1}{Re} \nabla^2 \mathbf{u},  \quad \text{in} \quad \Omega \times (0, T),
\end{equation}
\begin{equation}
    \nabla \cdot \mathbf{u} = 0, \quad \text{in} \quad \Omega \times (0, T),
\end{equation}
\end{subequations}
where $Re = \overline{U}D / \nu$ denotes the Reynolds number, with $\overline{U}$ being the characteristic velocity at the inlet, $D$ as the characteristic length scale, which is the cylinder diameter, and $\nu$ the kinematic viscosity of the fluid.
The flow studied here occurs at $Re = 100$.

The inlet boundary $\Gamma_\text{in}$ of the computational domain is defined by a parabolic velocity profile, where the velocity is highest at the centerline and decreases symmetrically toward the edges, reaching zero at the walls. 
An outflow boundary condition is imposed on the outlet $\Gamma_\text{out}$.
The upper and lower boundaries are no-slip walls, where the fluid velocity is zero.
The boundaries of the synthetic jets $\Gamma_{i}$ (where $i = 1, 2$) have a parabolic velocity profile. To ensure mass flow balance, the velocities satisfy $V_{\Gamma_1} = -V_{\Gamma_2}$. A positive value indicates that the synthetic jet is in blowing mode, while a negative value indicates suction mode.
The rest of the elliptical cylinder, excluding the synthetic jets, has a no-slip boundary condition.
Mathematically, the boundary conditions are expressed as:
\begin{eqnarray}
\begin{array}{cl}
- p \mathbf{n} + Re^{-1} ( \nabla \mathbf{u} \cdot \mathbf{n}) = 0  &\text{on\quad } \;\Gamma_{out}, \\
\mathbf{u} = 0  &\text{on \quad} \Gamma_{w}, \\
\mathbf{u} = U  &\text{on \quad} \Gamma_{in}, \\
\mathbf{u} = f_{Q_i}  & \text{on \quad} \Gamma_{i}, \quad i = 1, 2.
\end{array}
\end{eqnarray}
where $\mathbf{n}$ denotes the unit vector normal to the outlet, $U$ is the inflow velocity profile, and $f_{Q_i}$ represents the velocity profiles for the jets, simulating fluid suction or injection.
More detailed descriptions related to the numerical method settings are provided in Appendix A.

The simulations are conducted using the open-source \texttt{OpenFOAM} software, which utilizes the finite volume method to discretize the computational domain \cite{jasak2009}.
A hybrid mesh generation method is employed to ensure high accuracy around the cylinder while minimizing computational time.
The near-wall region of the elliptical cylinder is discretized with multiple layers of quadrilateral meshes. This approach enhances computational accuracy and aligns more closely with the flow dynamics. Meanwhile, the rest of the domain employs triangular meshes to efficiently manage complex geometries and transitions.
The discretization results for the elliptical cylinder and  the mesh independency tests are shown in Appendix B.

We use the \texttt{pimpleFoam} solver to solve incompressible fluid flow problems.
This solver combines the PISO and SIMPLE algorithms \cite{ISSA198640,SIMPLE}.
The hybrid approach improves numerical stability and convergence for steady and transient incompressible flow simulations, leading to higher computational accuracy.
The lift coefficient $C_L$ and drag coefficient $C_D$ are key parameters characterizing the aerodynamic performance of the cylinder, where $C_L$ represents the force perpendicular to the flow and $C_D$ quantifies the flow resistance.
These coefficients are defined as:
\begin{align}
     C_L = \frac{F_L}{0.5\rho \overline{U}^2D}, \quad  C_D = \frac{F_D}{0.5\rho \overline{U}^2D},
\end{align}
where $F_L$ and $F_D$ represent the lift and drag forces, respectively, which are integrated over the surface of the elliptical cylinder.  

\subsection{Deep reinforcement learning}\label{sec:ANN}

The fundamental DRL framework comprises two main components: the agent and the environment. The agent, typically an artificial neural network trained via a trial-and-error algorithm, interacts with the environment through three key elements.
(\romannumeral1) State observations $s_i$, which may be noisy and stochastic.
(\romannumeral2) Actions $a_i$, the control inputs to the environment.
(\romannumeral3) Reward signals $r_i$, which reflect the desirability of the current state.
This closed-loop interaction enables the agent to make decisions aimed at maximizing cumulative rewards by effectively interpreting state information.

\begin{figure*}[htb!]
    \centering
    \includegraphics{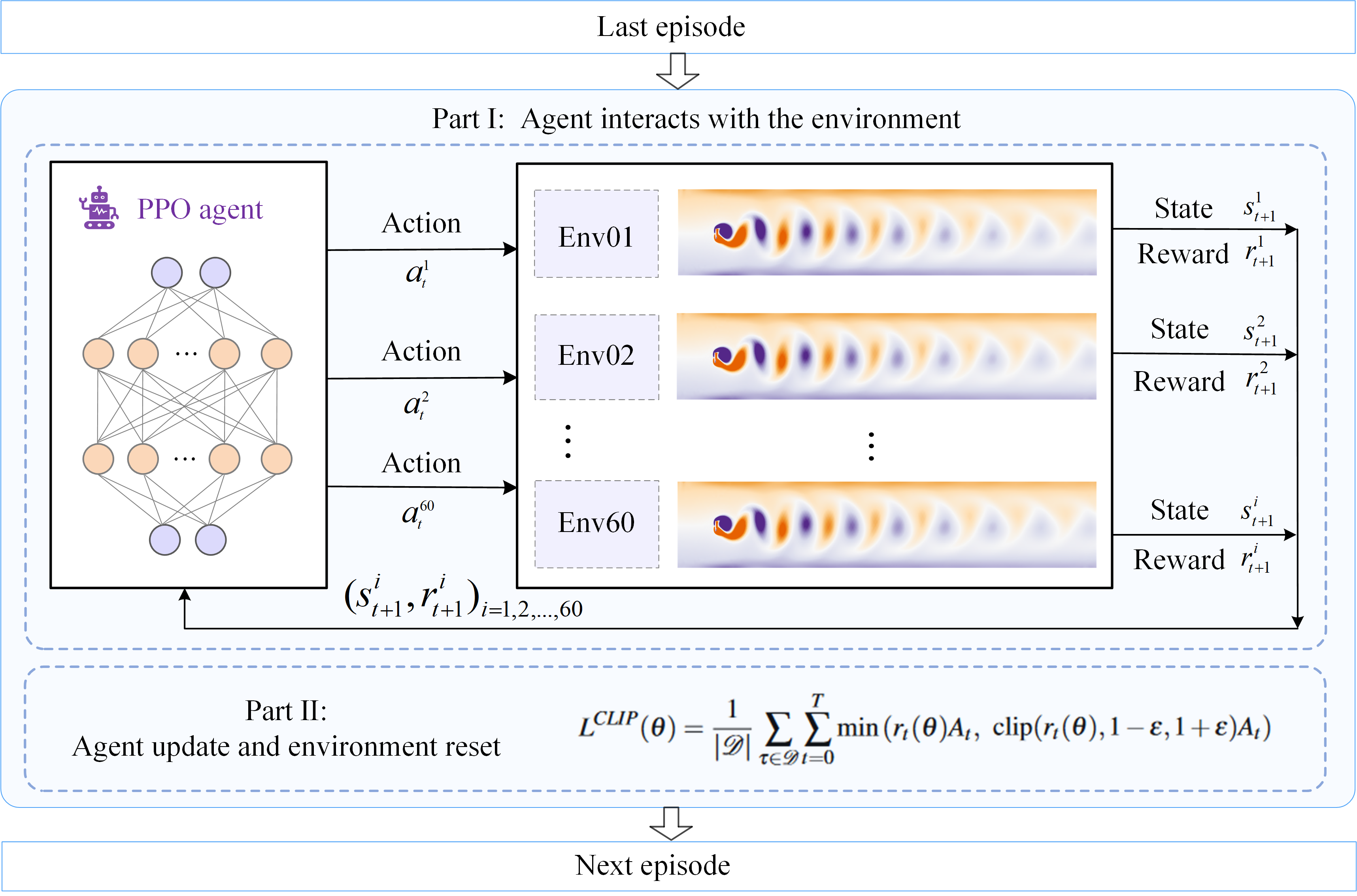}
    \caption{Schematic of the PPO agent interacting with multiple environments. Part I illustrates the parallel interaction process, where the agent simultaneously interacts with 60 independent environments. At each time step $t$, the agent receives the current states $s^i_t$, returns actions $a^i_t$, and collects the corresponding rewards $r^i_t$ from each environment. Part II shows the policy update stage. The collected trajectories are used to compute gradients and optimize the PPO objective function $L^{CLIP}{(\theta)}$. After each episode, all environments are reset for the next round of interaction. This setup enables efficient data collection and stable training in high-dimensional, multi-environment reinforcement learning tasks.}
    \label{fig:figure2}
\end{figure*}

The agent, powered by a neural network, observes the current state $s_i$ and generates an action $a_i$ through its policy network $\pi_{\theta}(a_i | s_i)$. This action is executed, leading to a state transition and a reward $r_i$, which is used to update the policy. DRL methods are broadly divided into policy-based and value-based approaches. Value-based methods, such as $Q$-learning, select actions by maximizing the Q-value function:
$a^* = \arg\max_{a} Q(s, a),$ where $a^*$ is the optimal action, and $Q(s, a)$ estimates the expected reward for taking action $a$ in state $s$. The Q-value function is updated using the Bellman equation:
\begin{equation}
Q(s, a) \leftarrow Q(s, a) + \alpha \left[ r + \gamma \max_{a'} Q(s', a') - Q(s, a) \right],
\end{equation}
where $\alpha$ is the learning rate, $r$ is the reward, $\gamma$ is the discount factor, and $s'$ is the next state. In contrast, policy-based methods optimize the policy parameters $\theta$ directly by maximizing the expected return. The policy gradient is given by:
\begin{equation}
\nabla_{\theta} J(\theta) = \mathbb{E}_{\tau \sim \pi_{\theta}} \left[ \nabla_{\theta} \log \pi_{\theta}(a|s) \cdot \hat{R}(\tau) \right],
\end{equation}
where $ \nabla_{\theta} J(\theta) $ adjusts policy parameters $ \theta $ to maximize expected returns, with $ \mathbb{E}_{\tau \sim \pi_{\theta}} $ denoting the expectation over trajectories sampled from $ \pi_{\theta} $. 
The term $ \nabla_{\theta} \log \pi_{\theta}(a|s) $ quantifies the sensitivity of action probabilities to parameter changes, and $ \hat{R}(\tau) $ reflects the cumulative return from the trajectory.

PPO is an advanced reinforcement learning algorithm that is particularly well-suited for solving continuous control problems. 
It is a model-free, on-policy algorithm that optimizes policy updates using a clipped objective function, ensuring stable and efficient learning. One of the key advantages of PPO is its simplicity and reliability in handling high-dimensional and continuous action spaces, which makes it ideal for applications like flow control where the actions (such as jet mass flow rates) are continuous. We describe the process of using the PPO algorithm to interact with numerical simulation in \Cref{fig:figure2}, which involves the generation of ($s_i$, $a_i$, $r_i$) training data and the improvement of the control strategy.
The objective function formula, parameter update method and iteration process of the PPO algorithm are shown in Appendix C.

\begin{figure*}
    \centering
    \includegraphics{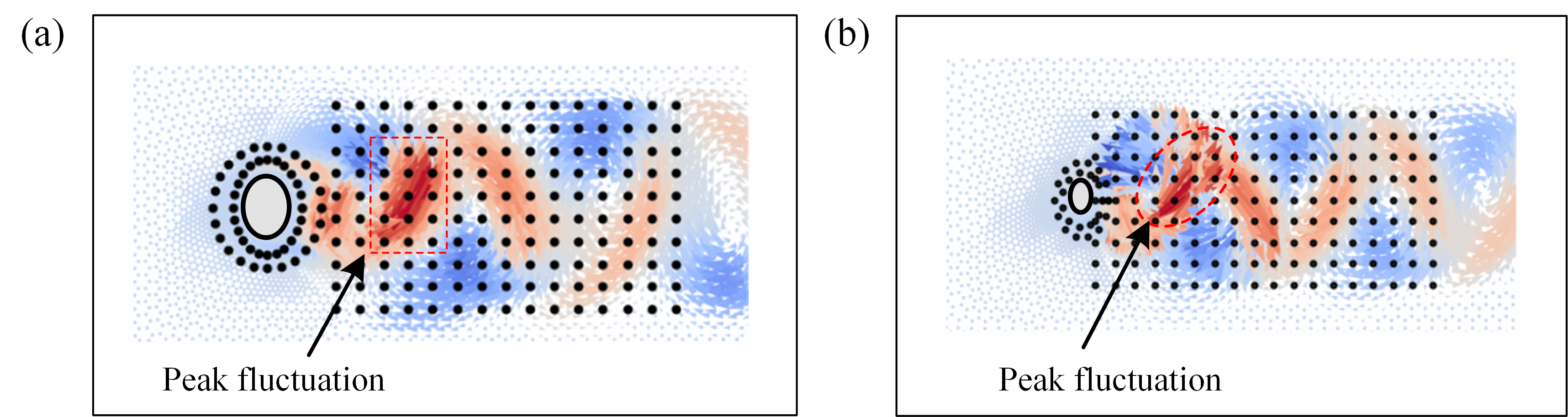}
    \caption{Instantaneous velocity field of the flow past an elliptical cylinder with $Ar = 0.75$ and different $\beta$, highlighting the overlap between flow fluctuation and probe distribution.
    (a) Velocity field for $\beta = 0.24$, showing strong vortex shedding in the near-wake region. 
    (b) Velocity field for $\beta = 0.12$, showing broader and more dispersed wake region.
    In both cases, the probes are positioned in regions with significant fluctuation, ensuring representative and sensitive measurements for the subsequent evaluation of control performance.}
    \label{fig:figure3}
\end{figure*}

\subsection{DRL-based active flow control}\label{sec:DRL_AFC}

We employ an intelligent interaction framework that integrates DRL with CFD to address the problem of AFC. Within this framework, DRL conceptualizes the fluid dynamics simulation as an interactive environment, with three primary channels of interaction. Observations are represented by $s_t$, which involve extracting measurements at probe points from CFD simulations. Actions are denoted by $a_t$, corresponding to the modulation of the mass flow rates of two synthetic jets positioned on the cylinder. 
The reward function $r_t$ is designed to minimize the drag coefficient while introducing a penalization term based on the lift coefficient to discourage unsteady or asymmetric control behaviors.

\begin{itemize}
\item State space $ \mathcal{S}.$ In the flow field around the cylinder, regions with the highest fluctuation values typically indicate areas of pronounced instability, characterized by significant velocity or pressure fluctuations caused by shear effects or vortex shedding. As shown in \Cref{fig:figure3}, the red arrow highlights the peak fluctuation region, marking a critical area in the flow dynamics. To capture these variations effectively, we strategically placed observation probes both around the elliptical cylinder and within the regions exhibiting the most intense fluctuation. The velocity or pressure measurements from these probes are then provided to the agent as the environmental state. The optimization of probe placement and quantity can be guided by \cite{parisRobust, liReinforcementlearning}, while additional insights into dynamic characteristics can be drawn from \cite{Xia2024, Dynamic2024}.

\item Action space $ \mathcal{A}.$ The control action is defined by the dimensionless mass flow rates $Q_1$ and $Q_2$ of synthetic jets on the cylinder, constrained by $Q_1 + Q_2 = 0$ to ensure zero net mass flow. 
The function $S(V_{\Gamma_1,T_i}, a, V_{\Gamma_1,T_{i-1}})$, designed to facilitate a smooth transition from the current value $V_{\Gamma_1,T_i}$ to the target action $a$, is defined as:
\begin{equation}
S(V_{\Gamma_1,T_i}, a, V_{\Gamma_1,T_{i-1}}) = V_{\Gamma_1,T_i} + \alpha \cdot (a - V_{\Gamma_1,T_{i-1}}),
\end{equation}
where $V_{\Gamma_1,T_i}$ represents the current state at time step $i$, $a$ is the target action, and $V_{\Gamma_1,T_{i-1}}$ is the previous state. The $\alpha \in (0, 1]$ is a relaxation factor that controls the rate of adjustment, thereby preventing abrupt changes in actuation and enhancing numerical stability.

\item Reward mechanism $ r_t $. Drawing inspiration from the work of Rabault et al. \cite{rabault2019artificial}, the reward function utilized in the present is designed to incentivize actions that reduce drag and lift coefficient.
The reward function is defined as:
\begin{equation}
r_t = (\langle C_D \rangle_{T_0} - \langle C_D \rangle_T) - 0.1 \left|\langle C_L \rangle_T\right|,
\end{equation}
where, $\langle C_D \rangle_{T_0}$ denotes the mean drag coefficient of the cylinder over a baseline period $T_0$ without any control measures. 
$\langle C_D \rangle_T$ represents the mean drag coefficient during the controlled period $T$.
The mean lift coefficient $\langle C_L \rangle_T$ over the controlled period is penalized by a factor of 0.1 to discourage control strategies that reduce drag at the cost of increased lift, which could destabilize the cylinder.
\end{itemize}

In our framework, the DRL network trains online with the CFD simulation without any prior training phase. 
After each simulation timestep, the updated flow state is immediately fed into the DRL network to optimize the control policy, ensuring that the agent continuously adapts to the changing flow characteristics in real-time. 
This integrated online learning strategy not only maintains computational efficiency but also improves the agent's adaptability to varying flow conditions.
By iteratively updating the policy within the simulation loop, the agent progressively develops more effective flow control strategies, achieving a balanced trade-off between real-time adaptation and computational performance.
The hyperparameters of the DRL framework employed in this study are presented in Appendix D.

\section{Results}\label{sec:results}

This section presents the application of DRL to flow control around elliptical cylinders. \Cref{sec:result01} reports the control results for an elliptical cylinder with $\beta = 0.24$, while \Cref{sec:result02} provides a detailed analysis of the partially controlled flow. \Cref{sec:result03} focuses on the case with a reduced $\beta$ of 0.12. These studies collectively investigate the scalability and optimization potential of DRL-based flow control strategies across elliptical cylinders with varying geometrical parameters.

\subsection{Elliptical cylinder with a $\beta$ of 0.24}\label{sec:result01}

This section investigates the impact of AFC using synthetic jets on a circular cylinder ($Ar = 1$) and elliptical cylinders ($Ar = 0.75, 0.5, 0.25, 0.1$) with a $\beta=0.24$. 
The analysis encompasses the evolution of the reward function during the DRL training process in \Cref{DRL training process}, the fluid force suppression effects in \Cref{fluid force suppression}, the evolution of flow structures under controlled conditions in \Cref{Flow structure01}, the suppression of wake vortices in \Cref{Eliminating vortices}, and the assessment of energy consumption associated with the flow control strategies in \Cref{Energy consumption}.

\subsubsection{DRL training process}\label{DRL training process}

\begin{figure*}[ht]
    \centering
    \includegraphics[width=\textwidth]{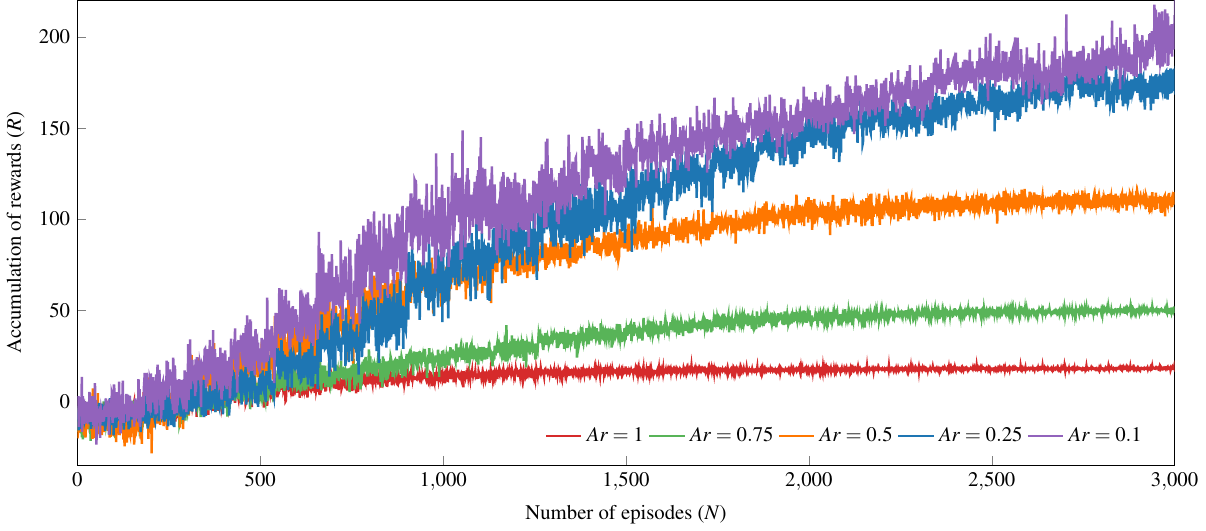}
    \caption{Training history of the PPO algorithm for active flow control around elliptical cylinders with aspect ratios $Ar = 1.0$, $0.75$, $0.5$, $0.25$, and $0.1$, at a fixed blockage ratio of $\beta = 0.24$.
    The figure shows that under different $Ar$, the training reward increases with the number of episodes, and a smaller $Ar$ leads to higher cumulative rewards.}
    \label{fig:figure4}
\end{figure*}

In all training setups, the fully developed baseline flow serves as the initial state for flow control. Flow control is then applied based on this baseline flow. The DRL learning curves for the cylinder with an aspect ratio of $Ar = 1$ and for elliptical cylinders with aspect ratios of $Ar = 0.75, 0.5, 0.25, 0.1$ are depicted in \Cref{fig:figure4}.
Each DRL training session involves running 3000 episodes, with 100 timesteps per episode.
The reward at each timestep reflects the effectiveness of the action taken, and the total reward accumulated over an episode is called the episode return.
For the cylinder with $Ar = 1$, the DRL learning curve stabilizes after approximately 1,000 episodes, indicating that the episode return has reached a consistent level.
In contrast, for elliptical cylinders with $Ar = 0.75$ and $Ar = 0.5$, the cumulative reward initially increases before peaking, reflecting a slower convergence. As the aspect ratio $Ar$ decreases, the stability of the DRL training return becomes significantly compromised.
When the $Ar$ of the elliptical cylinder is 0.25 and 0.1, the training speed and stability of the reward function trained by DRL significantly decrease compared to cases with an $Ar$ greater than 0.25.
This decline results from the slender geometry of these elliptical cylinders, which intensifies flow instability. Consequently, achieving effective flow control becomes increasingly challenging, negatively impacting the stability and convergence of the DRL training process.

As the aspect ratio of the elliptical cylinder decreases, the rate of increase in the reward function during DRL training gradually reduces. This indicates that the difficulty of flow control increases as the cylinder shape evolves from circular to slender elliptical. Sharper edges of the ellipse induce stronger flow separation, which poses greater challenges to flow control. This phenomenon resembles the additional difficulties encountered in controlling vortex shedding around square cylinders compared to circular ones. Similar observations are reported in Yan et al. \cite{yan2023stabilizing} and Jia et al. \cite{jia2024jetsactuator}, where enhanced separation at sharp edges leads to increased control complexity. Considering the correlation between separation intensity and flow control effectiveness, further investigation into the relationship between edge sharpness and the robustness of developed control strategies proves valuable.

\subsubsection{Suppress fluid forces}\label{fluid force suppression}

\begin{figure*}[hbt!]
    \centering
    \includegraphics[width=\textwidth]{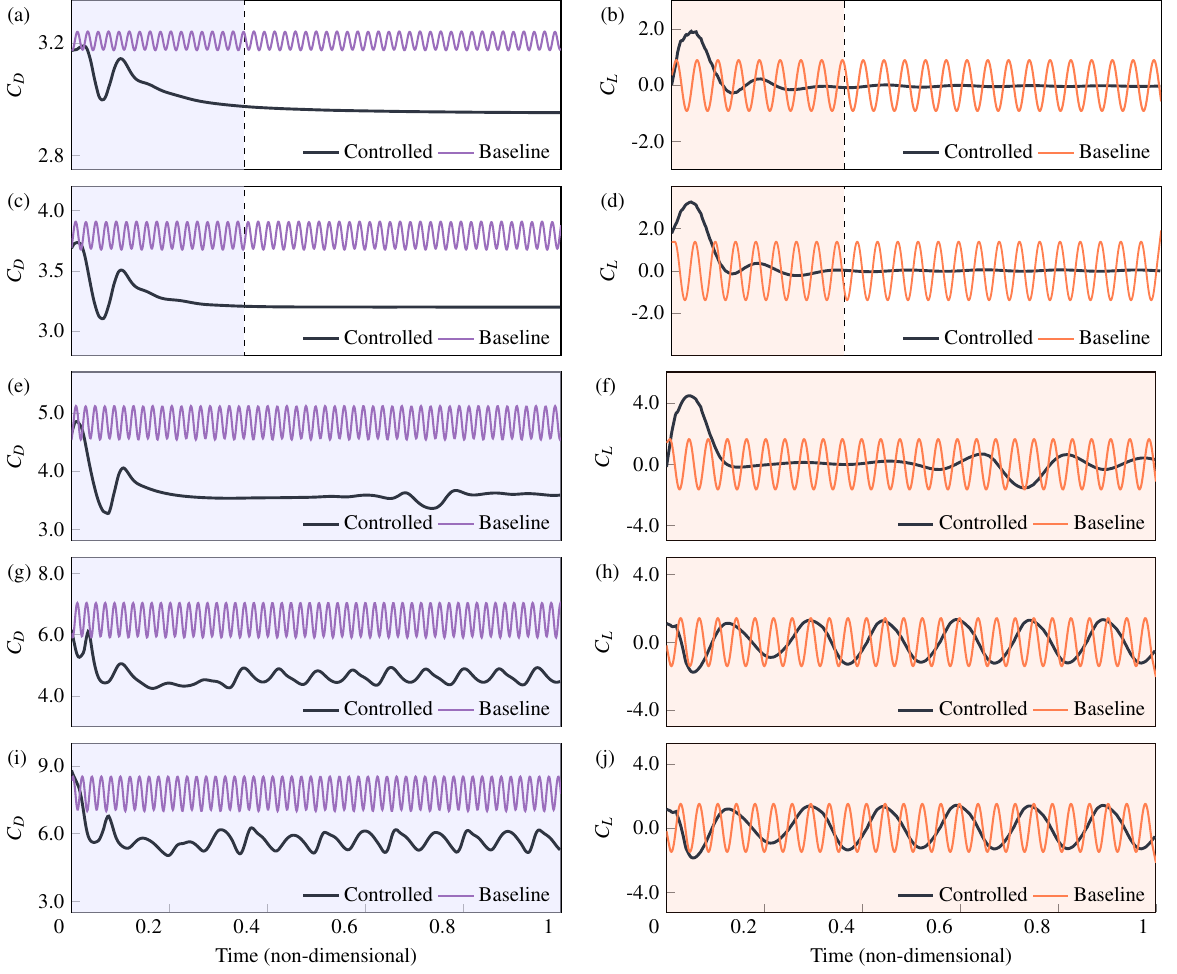}
    \caption{Time histories of the drag coefficient $C_D$ and lift coefficient $C_L$ before and after applying DRL-based flow control for elliptical cylinders with $Ar$ ranging from 1.0 to 0.1 and a $\beta$ of 0.24. 
    The left column shows the temporal evolution of $C_D$, while the right column shows $C_L$ for each $Ar$.
    Colored backgrounds indicate periods of unsteadiness, while white backgrounds denote steady-state behavior after control. 
    Each pair of subfigures compares the baseline and controlled cases for a specific $Ar$ value:
    (a) $C_D$ for $Ar = 1.0$; (b) $C_L$ for $Ar = 1.0$; (c) $C_D$ for $Ar = 0.75$; (d) $C_L$ for $Ar = 0.75$; (e) $C_D$ for $Ar = 0.5$; (f) $C_L$ for $Ar = 0.5$; (g) $C_D$ for $Ar = 0.25$; (h) $C_L$ for $Ar = 0.25$; (i) $C_D$ for $Ar = 0.1$; (j) $C_L$ for $Ar = 0.1$.}
    \label{fig:figure5}
\end{figure*}

The drag reduction effects of DRL-based flow control for cylinders with aspect ratios ranging from $Ar = 1$ to $Ar = 0.1$ are illustrated in \Cref{fig:figure5}.
For the cylinder with $Ar = 1$, the activation of synthetic jet actuators induces a rapid response in the drag coefficient, which immediately decreases to a minimum value before stabilizing near this level. The lift coefficient exhibits a similar trend, displaying a transient response upon the commencement of control and gradually diminishing to zero. The initially periodic oscillations in both drag and lift coefficients within the cylinder’s wake are effectively attenuated, leading to a stable flow state. 
For the elliptical cylinder with an $Ar$ of 1 at a $Re$ of 100, the controlled drag and lift coefficients obtained in this study demonstrate significantly smoother profiles compared to those reported by Varela et al. \cite{Varela2022}. In contrast, the control strategy proposed by Varela et al. \cite{Varela2022} in fluid force coefficients that persist in periodic oscillation, underscoring the enhanced stability and effectiveness of the control strategy developed in the present work.

The elliptical cylinder with $Ar = 0.75$ demonstrates a comparable drag reduction behavior under synthetic jet control, closely paralleling the performance observed for $Ar = 1$.
As the $Ar$ is further reduced to 0.5, the behavior of the lift and drag coefficients begins to diverge from the patterns observed at higher aspect ratios. While these coefficients initially exhibit a sharp decrease and subsequent stabilization upon the initiation of control, fluctuations become evident during the later stages of control. In the cases of elliptical cylinders with $Ar = 0.25$ and 0.1, the reduction in lift and drag coefficients displays marked deviations from the trends observed at higher $Ar$. 
Following an initial gradual decline, these coefficients settle into a state of periodic oscillation. Compared to the baseline flow, the amplitude and frequency of these oscillations are reduced, suggesting that flow instability is partially mitigated under control for the elliptical cylinders with $Ar = 0.25$ and 0.1.

\subsubsection{Flow structure comparison}\label{Flow structure01}

\begin{figure*}
    \centering
    \includegraphics{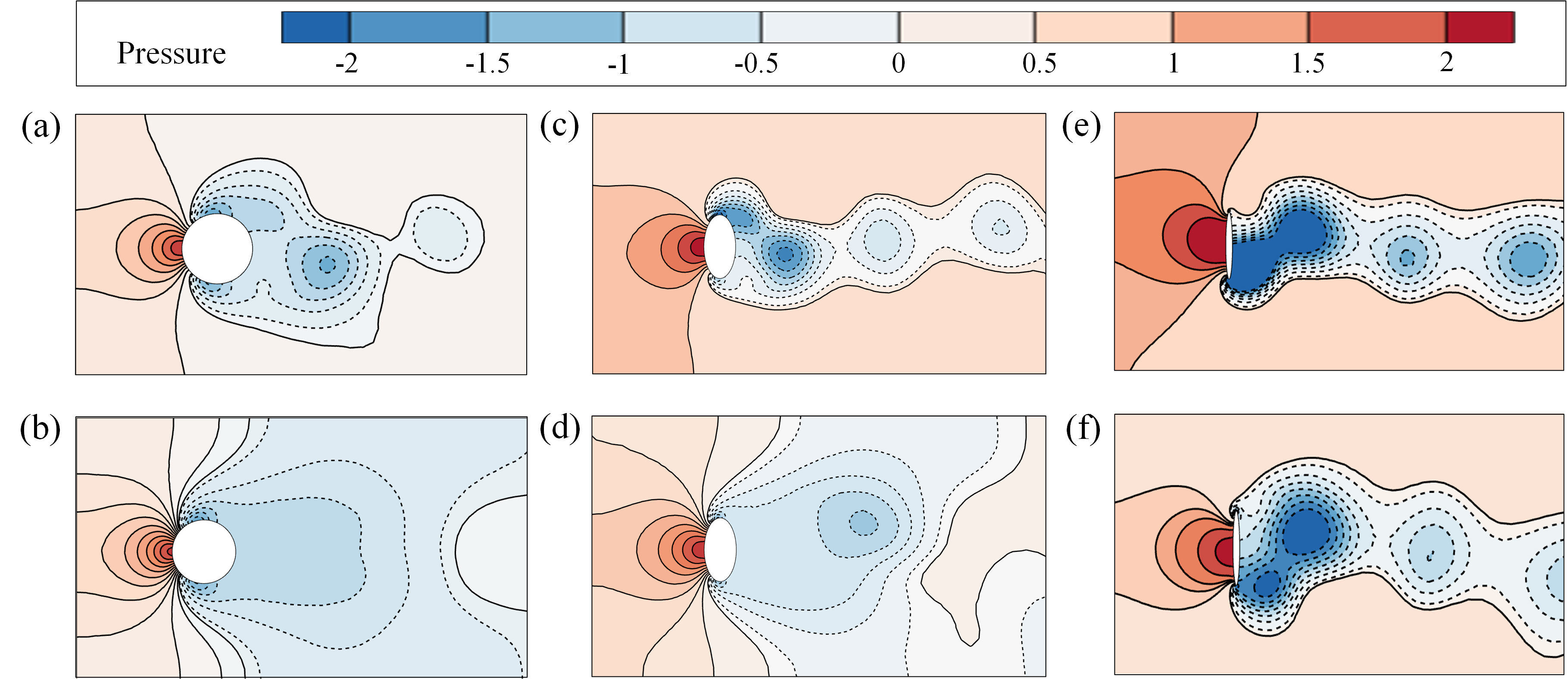}
    \caption{Instantaneous pressure contours of the flow field before (top row) and after (bottom row) active control for elliptical cylinders with three representative aspect ratios. 
    The figure illustrates that active flow control leads to a more uniform and symmetric pressure distribution around elliptical cylinders across different aspect ratios.
    (a) Baseline flow for $Ar$ = 1; (b) Controlled flow for $Ar$ = 1; 
    (c) Baseline flow for $Ar$ = 0.5; (d) Controlled flow for $Ar$ = 0.5; 
    (e) Baseline flow for $Ar$ = 0.1; (f) Controlled flow for $Ar$ = 0.1.}
    \label{fig:figure6}
\end{figure*}

The instantaneous pressure contours around the cylinder, before and after the implementation of flow control, are illustrated in \Cref{fig:figure6}.
To concisely capture the key phenomena, we present cases with aspect ratios $Ar$ = 1, 0.5, and 0.1. 
For clarity, zero-pressure contour lines are highlighted to emphasize the impact of flow control on the instantaneous pressure distribution around the elliptical cylinders.
For the cylinder with $Ar = 1$, the baseline flow reveals the formation of an asymmetric low-pressure region behind the cylinder. Under controlled flow conditions, the instantaneous pressure contours become symmetric, with a significant expansion of the low-pressure region around the cylinder. Additionally, the peak negative pressure is markedly reduced compared to that observed in the baseline flow.
This indicates that the flow control strategy effectively modifies the pressure distribution, leading to more stable and balanced flow characteristics.

For the elliptical cylinder with $Ar = 0.5$, the baseline flow exhibits an instantaneous pressure distribution around the cylinder similar to that observed for the cylinder with $Ar = 1$. 
Compared to the baseline flow, the instantaneous pressure contours under controlled flow conditions become milder, characterized by an expanded low-pressure region and a reduction in the peak negative pressure.
Under controlled conditions, the instantaneous pressure distribution of the elliptical cylinder with $Ar = 0.5$ is significantly less symmetric and uniform compared to that of the cylinder with $Ar = 1$.
When the $Ar$ is further reduced to 0.1, the change in the instantaneous pressure contours before and after control becomes less significant, with only the expansion of the low-pressure region and a reduction in the peak negative pressure observed. 

Comparing the control performance across the three elliptical cylinders, it is evident that as the aspect ratio decreases, the ability to achieve symmetry and uniformity in the pressure distribution diminishes. The control strategy is more effective for cylinders with higher aspect ratios, where the pressure distribution can be more symmetrically adjusted. In contrast, for extremely low aspect ratios, such as $Ar = 0.1$, the control effect becomes less pronounced, indicating that flow control becomes more challenging as the geometry of the cylinder becomes increasingly slender.

\begin{figure*}
    \centering
    \includegraphics{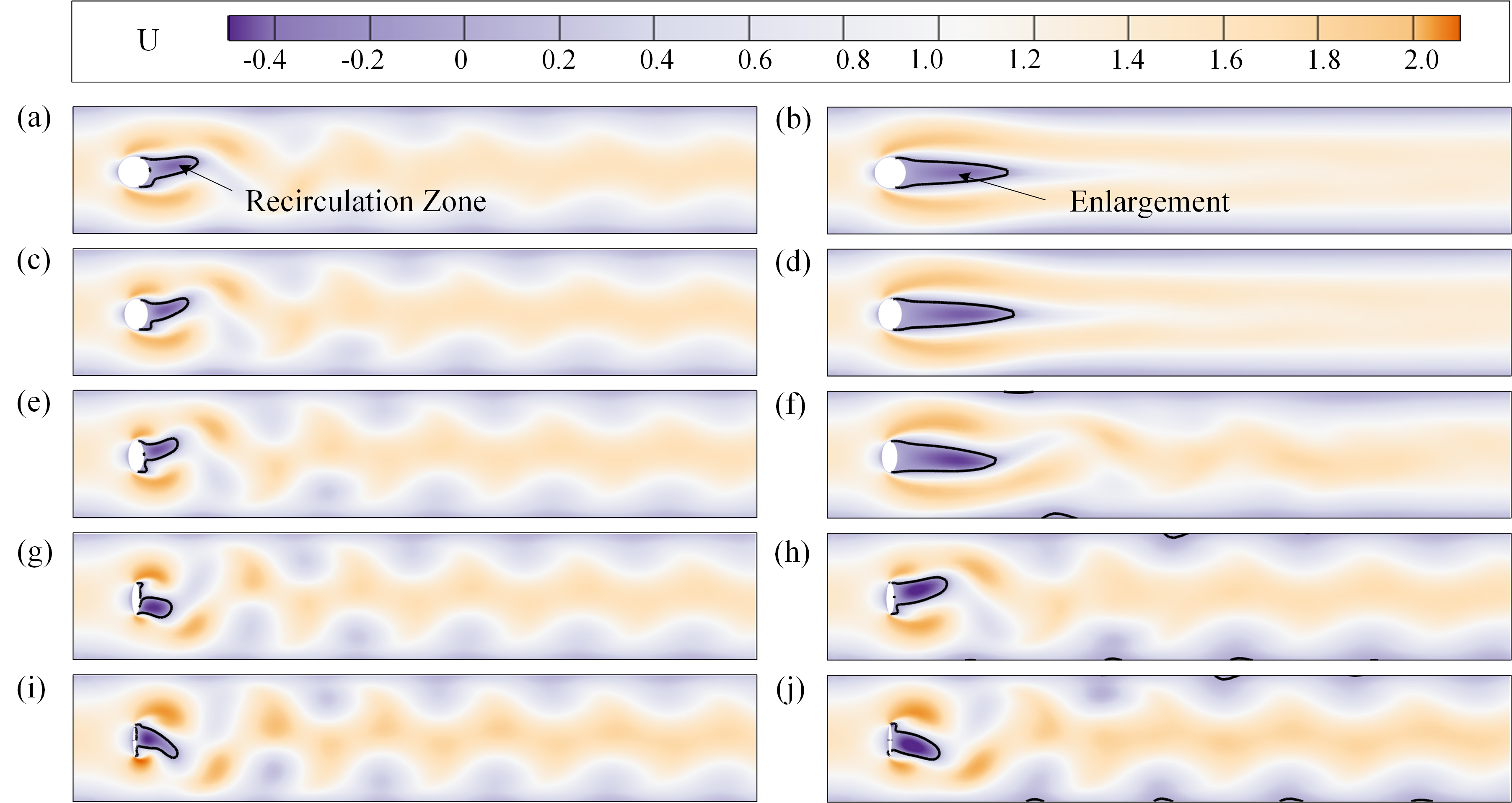}
    \caption{Instantaneous velocity contours before (left) and after (right) active flow control for elliptical cylinders with varying aspect ratios. 
    For each aspect ratio, the left panel shows the baseline flow, while the right panel illustrates the controlled flow modified by synthetic jet actuation.
    (a) Baseline velocity field for $Ar = 1.0$. (b) Controlled flow for $Ar = 1.0$. 
    (c) Baseline flow for $Ar = 0.75$. (d) Controlled flow for $Ar = 0.75$. 
    (e) Baseline flow for $Ar = 0.5$. (f) Controlled flow for $Ar = 0.5$. 
    (g) Baseline flow for $Ar = 0.25$. (h) Controlled flow for $Ar = 0.25$. 
    (i) Baseline flow for $Ar = 0.1$. (j) Controlled flow for $Ar = 0.1$.}
    \label{fig:figure7}
\end{figure*}

The instantaneous velocity contours around the cylinder, before and after the implementation of flow control, are illustrated in \Cref{fig:figure7}. To visually emphasize the recirculation regions, zero-velocity isocontours are included in the figure.
We define the recirculation area as the region in the downstream neighborhood of the cylinder where the horizontal component of velocity is negative.
For the cylinder with $Ar = 1$, a distinct recirculation region forms behind the cylinder in the baseline flow. After flow control is applied, the size of the recirculation region behind the cylinder increases significantly, and the overall velocity distribution becomes more symmetric. For the elliptical cylinder with $Ar = 0.75$, the patterns observed in the instantaneous velocity contours are generally consistent with those of the circular cylinder, including a similar expansion of the recirculation region and an enhancement in the symmetry of the velocity distribution.
This phenomenon that the range of the recirculation area in the controlled flow is significantly increased is consistent with the results of Rabault et al. \cite{rabault2019artificial}, which illustrates the effectiveness of the control strategy in reducing the impact of vortex shedding.

For the elliptical cylinder with $Ar = 0.5$, the recirculation region behind the elliptical cylinder elongates significantly compared to the baseline flow. 
However, in contrast to the cases with $Ar = 0.75$ and $Ar = 1$, the velocity distribution in the wake under control exhibits more intense fluctuations.
For elliptical cylinders with $Ar = 0.25$ and $Ar = 0.1$, the recirculation region in the controlled flow only exhibits slight expansion, with no significant elongation. The velocity distribution in the wake region shows minimal differences between the baseline and controlled flows.
It is clear that flow control is most effective for the cylinders with higher aspect ratios, where the recirculation region expands significantly, and the velocity distribution becomes more symmetric. 
For cylinders with lower $Ar$, especially those with $Ar = 0.25$ and $Ar = 0.1$, the control effects are less pronounced, with minimal changes in the recirculation region and velocity distribution, indicating that the control strategy becomes less effective as the aspect ratio decreases.

\subsubsection{Eliminating vortices}\label{Eliminating vortices}

The vorticity distribution around the cylinder reflects the characteristics of the wake structure and flow separation.
Vorticity is calculated as the curl of the velocity field and can be expressed on the cylinder surface as the velocity gradient normal to the surface.
To clearly illustrate the changes in vortex shedding in the wake of the elliptical cylinders, we present instantaneous vorticity contours for $Ar = 1$, 0.75, 0.5, 0.25, and 0.1. 
The close-up view of the vorticity field for cylinders with different $Ar$ under baseline flow conditions reveals that the vorticity field exhibits antisymmetry along the wake centerline. Taking the case of a cylinder with $Ar = 1$ as an example, the vorticity on the upper side of the centerline remains negative, except in a small region near the rear surface of the cylinder, where it becomes positive. This phenomenon, known as ``penetrated vorticity'', has been used by Mishra et al. \cite{MISHRA2021135} to explain the growth of the wake. The presence of penetrated vorticity causes the shear layers on both sides of the cylinder to diverge outward. 
As $Ar$ decreases, the penetrated vorticity intensifies, leading to increased separation between the shear layers and a wider wake region.
The phenomenon observed as $Ar$ decreases in the baseline flow of this study is consistent with the findings of Kumar et al. \cite{kumar2022steady}.

\begin{figure*}[hbt!]
    \centering
    \includegraphics{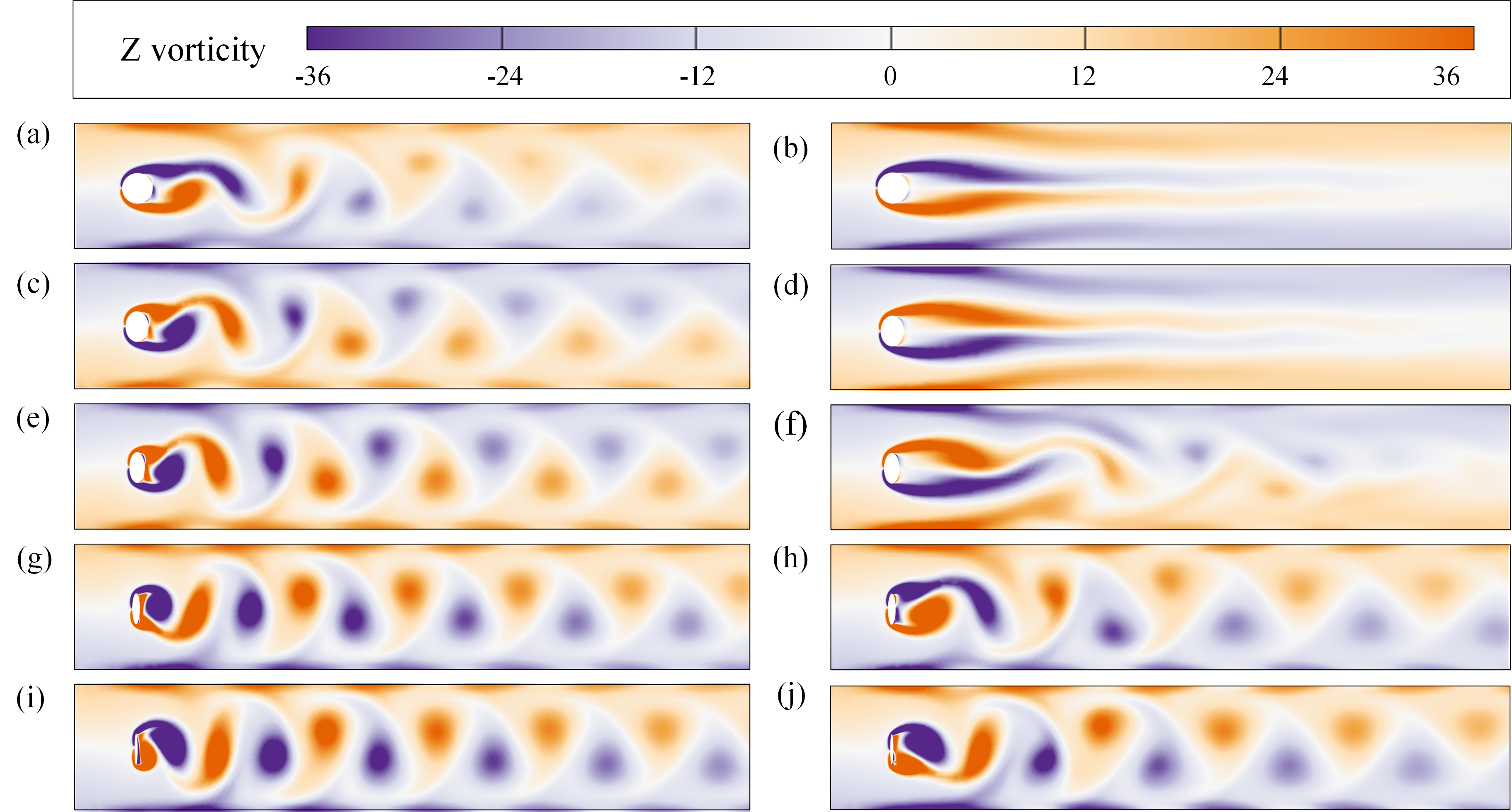}
    \caption{Instantaneous vorticity contours before (left) and after (right) active flow control for elliptical cylinders with varying aspect ratios. The left column in each pair represents the baseline flow, while the right column shows the controlled flow achieved through synthetic jet actuation. (a) Baseline flow for $Ar = 1.0$. (b) Controlled flow for $Ar = 1.0$. (c) Baseline flow for $Ar = 0.75$. (d) Controlled flow for $Ar = 0.75$. (e) Baseline flow for $Ar = 0.5$. (f) Controlled flow for $Ar = 0.5$. (g) Baseline flow for $Ar = 0.25$. (h) Controlled flow for $Ar = 0.25$. (i) Baseline flow for $Ar = 0.1$. (j) Controlled flow for $Ar = 0.1$.}
    \label{fig:figure8}
\end{figure*}

\Cref{fig:figure8} presents the instantaneous vorticity contours of the elliptical cylinders subjected to flow control. For elliptical cylinders with aspect ratios $Ar = 1$ and $Ar = 0.75$, the controlled flow effectively eliminates the periodic vortex shedding observed in the baseline cases, demonstrating complete suppression of wake unsteadiness. In contrast, for the case $Ar = 0.5$, despite the flow control strategy targeting a comparable degree of vortex suppression, the control efficacy diminishes, as evidenced by the persistence of elongated vortical structures in the wake. This partial suppression highlights the increasing complexity of flow dynamics as the cylinder becomes more slender. For even more slender elliptical cylinders, with $Ar = 0.25$ and $Ar = 0.1$, pronounced periodic vortex shedding remains clearly observable before and after control implementation. This persistence underscores the intrinsic difficulty in achieving full vortex shedding suppression in highly slender geometries, suggesting that conventional flow control strategies may require further adaptation or enhancement to address the amplified wake instabilities associated with such $Ar$.

Observations of the baseline flow around elliptical cylinders with aspect ratios ranging from 1 to 0.1 reveal that as the aspect ratio decreases, the length of the separation bubble progressively shortens. This reduction indicates that the flow separation becomes more pronounced as the elliptical cylinder becomes more slender, leading to increased control difficulty. In particular, elliptical cylinders with small aspect ratios exhibit stronger separation and intensified wake dynamics, similar to the challenges encountered when controlling vortex shedding around square cylinders compared to circular ones. 
This phenomenon aligns with the findings of Yan et al. \cite{yan2023stabilizing} and Jia et al. \cite{jia2024jetsactuator}. 
From a control perspective, the sharp-edged geometry of elliptical cylinders with low aspect ratios introduces additional instability modes, requiring the control system to respond to a broader spectrum of disturbances. 
Further research should therefore focus on quantifying the relationship between edge sharpness and flow separation strength, particularly in the context of energy-efficient control strategies.

\subsubsection{Energy consumption}\label{Energy consumption}

The primary objective of flow control is to actively manipulate fluid behavior in order to reduce drag, suppress vortex shedding, and improve overall flow characteristics. Among these, minimizing external energy input is critical, as energy consumption is directly tied to the system’s operational efficiency. Efficient control strategies must therefore strike a balance between aerodynamic performance and energy expenditure. However, achieving this balance becomes particularly challenging in flows around slender elliptical cylinders, where inherent flow instabilities demand more intensive control effort.

To quantify this trade-off, the control performance and associated energy costs for elliptical cylinders with various aspect ratios are summarized in \Cref{tab:table2}. 
Here, $\overline{C}_{D,\text{Baseline}}$ denotes the time-averaged drag coefficient of the baseline flow, while $\overline{C}_{D,\text{Controlled}}$ corresponds to that of the controlled flow. The drag reduction rate, $R_D$, represents the relative decrease in drag achieved through control. Additionally, $\overline{C}_{L,\text{Controlled}}$ indicates the time-averaged lift coefficient under controlled conditions. The actuation intensity is characterized by $\overline{a}$, the mean actuation amplitude, and $\overline{a}$ ratio, defined as the time-averaged synthetic jet mass flow rate normalized by the incoming flow rate at the inlet. These results highlight the trade-off between control effectiveness and energy expenditure, which varies significantly across different $Ar$ configurations.

\begin{table*}[hbt!]
  \centering
  \caption{Control performance and actuation energy consumption for elliptical cylinders with varying aspect ratios at a fixed blockage ratio ($\beta = 0.24$).}
  \begin{tabularx}{\textwidth}{ 
    >{\centering\arraybackslash}p{0.075\linewidth}
    >{\centering\arraybackslash}p{0.075\linewidth}
    >{\centering\arraybackslash}p{0.11\linewidth}
    >{\centering\arraybackslash}p{0.11\linewidth}
    >{\centering\arraybackslash}p{0.11\linewidth}
    >{\centering\arraybackslash}p{0.11\linewidth}
    >{\centering\arraybackslash}p{0.09\linewidth}  
    >{\centering\arraybackslash}p{0.11\linewidth}     
  }
\toprule
    $Ar$ & $Re$ & $\overline{C}_{D,\text{Baseline}}$ & $\overline{C}_{D,\text{Controlled}}$ & $R_D$ (\%) &  $\overline{C}_{L,\text{Controlled}}$ & $\overline{a}$ & $\overline{a}$ ratio (\%) \\
\hline   
    1.00    & 100 & 3.207 & 2.953 & 7.9   & -0.041 & 0.001 & 0.1  \\
    0.75    & 100 & 3.792 & 3.198 & 15.7  & 0.005  & 0.010  & 1.0  \\
    0.50    & 100 & 4.837 & 3.568 & 26.2  & -0.208 & 0.049 & 4.9  \\
    0.25    & 100 & 6.503 & 4.619 & 29.0  & -0.051 & 0.126 & 12.6 \\
    0.10    & 100 & 7.812 & 5.702 & 27.0  & 0.153  & 0.145 & 14.5 \\
\bottomrule
\end{tabularx}
\label{tab:table2}
\end{table*}

For the cylinder with $Ar = 1$, the application of the DRL-based flow control strategy leads to a 7.9\% reduction in drag, while maintaining an average lift coefficient of -0.041. The external energy consumption is minimized, amounting to only 0.1\% of the incoming flow. Furthermore, the instantaneous vorticity analysis reveals that the periodic vortex shedding observed in the baseline flow is fully suppressed under the controlled conditions. This result effectively achieves nearly all flow control objectives, significantly reducing drag, eliminating vortex shedding, and preserving the stability of the system with minimal external energy input. 
Similarly, for the elliptical cylinder with $Ar = 0.75$, the DRL-based control strategy demonstrates performance nearly identical to that of the cylinder with $Ar = 1$, achieving a highly effective flow control solution. 
The results highlight the robustness of the DRL approach in optimizing flow dynamics and energy efficiency across varying geometries.

However, as the $Ar$ decreases further, although the drag reduction rate improves, the associated external energy consumption increases significantly. For $Ar = 0.5$, 0.25, and 0.1, the drag reduction rate reaches between 26\% and 29\%, but this improvement is accompanied by a gradual increase in the average lift coefficient and a substantial rise in the energy consumption ratio, which reaches up to 14.5\%. This control strategy is less effective, particularly when compared to the case of $Ar = 1$, where the energy consumption ratio is only 0.1\%. Specifically, for $Ar = 0.25$ and $Ar = 0.1$, the energy consumption ratio escalates to 12.6\% and 14.5\%, respectively.  
These results underscore the growing inefficiency of the control strategy as the cylinder's aspect ratio decreases, highlighting the challenge of balancing drag reduction with energy efficiency in slender geometries.

\subsection{Analysis of incomplete vortex suppression}\label{sec:result02}

\begin{figure*}[htb!]
    \centering
    \includegraphics{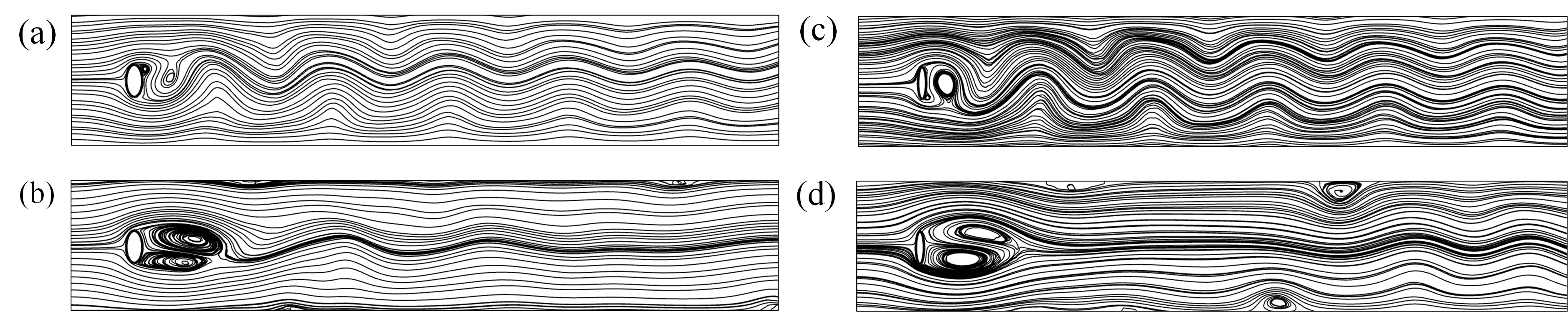}
    \caption{Comparison of streamline patterns between baseline and DRL-controlled flows around elliptical cylinders with aspect ratios $Ar = 0.5$ and $Ar = 0.25$.
    (a) $Ar$ = 0.5, baseline flow. (b) $Ar$ = 0.5, controlled flow.
    (c) $Ar$ = 0.25, baseline flow. (d) $Ar$ = 0.25, controlled flow.}
    \label{fig:figure9}
\end{figure*}

Comparative instantaneous vorticity contours before and after control for elliptical cylinders with aspect ratios $Ar=1$ and $Ar=0.75$ are presented to demonstrate the effectiveness of the flow control strategy in suppressing vortex shedding.
Under controlled flow conditions, the wake of the elliptical cylinder exhibits a stable and symmetric distribution of positive and negative vorticity, indicating that vortex shedding has been successfully mitigated.
However, as the $Ar$ decreases to 0.5, the suppression of vortex shedding becomes increasingly challenging. 
To further investigate the underlying mechanisms of this phenomenon, \Cref{fig:figure9} illustrates the streamline distributions before and after control implementation, providing deeper insights into the limitations of vortex shedding suppression.

A comparative analysis of the streamline distributions for elliptical cylinders with $Ar = 0.5$ and $Ar = 0.25$ under baseline and controlled flow conditions reveals significant differences in wake dynamics.
Under baseline flow conditions, the streamlines exhibit pronounced separation at the upstream region of the elliptical cylinder, leading to the formation of sizable vortex structures downstream.
This separation is a natural fluid dynamic response to the presence of an obstacle, resulting in unsteady vortices that increase drag. By contrast, under controlled flow conditions, the implementation of the jet control strategy effectively mitigates streamline separation, leading to a noticeable reduction in downstream vortex formation. 
Nevertheless, residual vortices near the sidewalls persist, suggesting that sidewall effects continue to influence the stability of the controlled flow. This finding indicates that although the control strategy demonstrates efficacy in suppressing vortex shedding, the sidewall-induced instabilities remain a critical factor impeding full stabilization.

For the elliptical cylinder with $Ar=0.25$, the baseline flow reveals even more pronounced separation at the upstream region, resulting in complex vortex structures in the wake. 
Under baseline flow conditions, the streamline separation at the upstream region is more pronounced, leading to the formation of increasingly complex vortex structures in the wake. This phenomenon arises due to the lower aspect ratio, which exacerbates flow complexity and intensifies fluid dynamic interactions around the obstacle. In contrast, under controlled flow conditions, the streamline distribution becomes noticeably smoother, and the formation of downstream vortices is partially suppressed.
Compared to an elliptical cylinder with an $Ar$ of 0.5, an elliptical cylinder with an $Ar$ of 0.25 exhibits more pronounced vortices in a controlled flow environment. On one hand, the lateral confinement enhances the interaction between coherent vortices and the side walls, which in turn reduces the effectiveness of control strategies in alleviating flow instability. On the other hand, the shape effects due to the lower aspect ratio also play a role. These findings suggest that the inherent flow instability is caused by the combined influence of side wall blockage and the geometric characteristics of the elliptical cylinder at low aspect ratios, both of which collectively impede the effectiveness of flow control strategies.

\begin{figure*}[hbt!]
    \centering
    \includegraphics{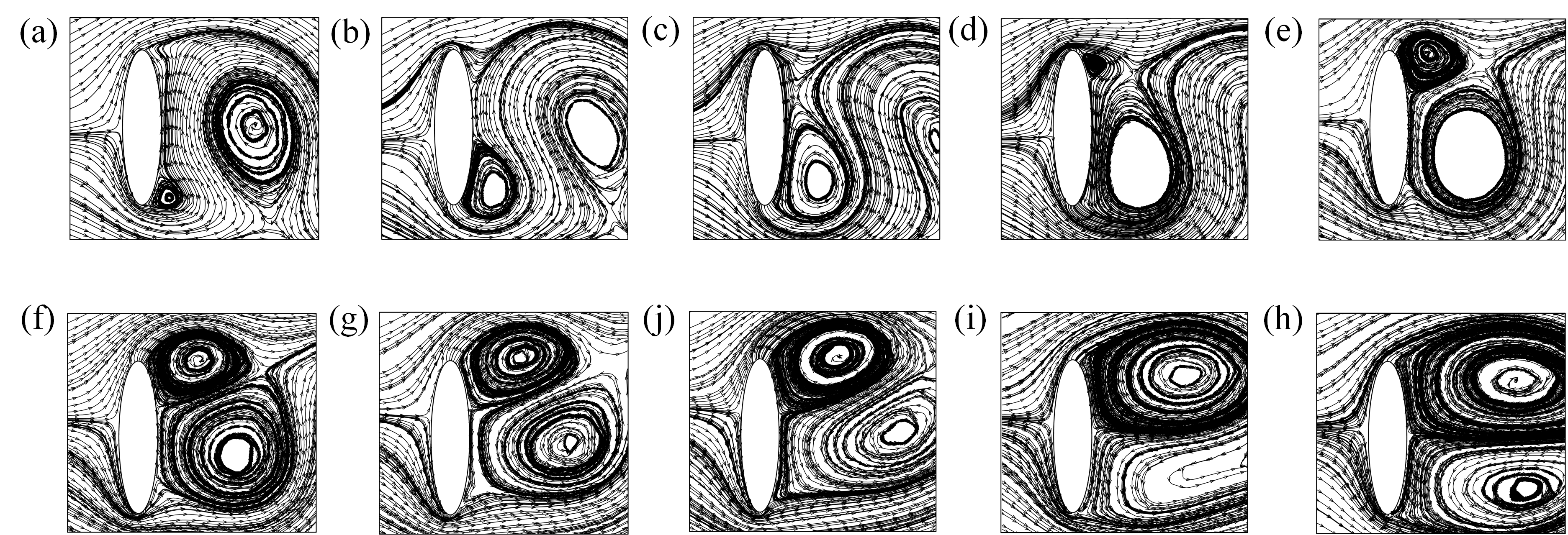}
    \caption{Streamline evolution around an elliptical cylinder with $Ar = 0.25$ during DRL-based flow control. Snapshots at successive time steps $t_0$–$t_9$ illustrate the transition from unsteady vortex shedding to a stabilized wake. (a) $t_0$: Initial asymmetric shedding; (b) $t_1$: Control onset disturbs wake structure; (c) $t_2$: Vortex interaction begins to change; (d) $t_3$: Enlarged recirculation region forms; (e) $t_4$: Strengthened vortex pairing; (f) $t_5$: Wake starts to reorganize; (g) $t_6$: Improved wake symmetry; (h) $t_7$: Further suppression of shedding; (i) $t_8$: Near-steady flow achieved; (j) $t_9$: Stable and symmetric wake maintained.}
    \label{fig:figure10}
\end{figure*}

The evolution of the flow field around the elliptical cylinder with $ Ar = 0.25 $ during the controlled cycle is illustrated in \Cref{fig:figure10}. 
This visualization captures the dynamics of vortex formation, shedding, and the subsequent stabilization of the wake.
Before the activation of the synthetic jets, the flow naturally separates from the surface of the elliptical cylinder, forming a primary vortex. As the synthetic jets are activated, the top and bottom jets induce localized high-speed fluid motion near the surface of the cylinder, disrupting the vortex formation process. Initially, small vortices form near the bottom of the cylinder, gradually growing into larger vortices, while a similar small vortex begins to form at the top. As both vortices at the top and bottom develop into larger vortices, they begin to compress toward each other. Eventually, these vortices coalesce and form a symmetric separation bubble at the rear of the cylinder. 
The flow control mechanism effectively suppresses vortex shedding, enhancing wake stability and transforming an unstable wake into a stable, symmetric recirculation bubble. 
These changes confirm the effectiveness of the active flow control strategy.

For an elliptical cylinder with a very low aspect ratio, a symmetrical separation bubble can still be formed on the back side of the elliptical cylinder in the controlled flow, which shows the effectiveness of the active flow control strategy.
The active flow control mechanism effectively suppresses vortex shedding, thereby enhancing wake stability and transforming an unstable wake into a stable, symmetric recirculation bubble. These results demonstrate the efficacy of the active flow control strategy applied to the elliptical cylinder. 
By examining the streamline distribution across the computational domain, a preliminary conclusion can be drawn: for highly slender elliptical cylinders, the instability of the controlled flow is governed primarily by the blockage ratio, rather than the aspect ratio.
In the following section, we will experimentally validate this conclusion.

\subsection{Elliptical cylinder with a $\beta$ of 0.12}\label{sec:result03}

In this section, the $\beta$ is set to 0.12 to systematically evaluate the effectiveness of active flow control in mitigating blockage effects. 
Five distinct elliptical cylinder configurations, characterized by $Ar$ of 1, 0.75, 0.5, 0.25, and 0.1, are examined. 
The flow control strategy employs DRL-based algorithms, with control actions executed via synthetic jet actuators. This approach enables a comprehensive assessment of the control efficacy across a range of geometric parameters.

\subsubsection{DRL training process}

\begin{figure*}[hbt!]
    \includegraphics[width=\textwidth]{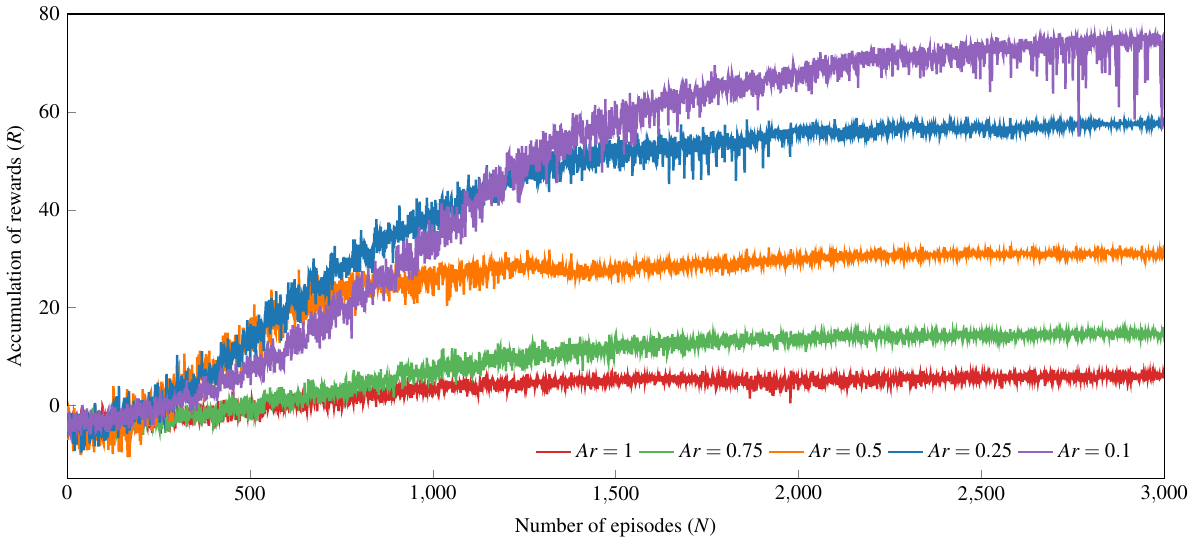}
    \caption{Training performance of deep reinforcement learning agents for flow control around elliptical cylinders with varying aspect ratios ($Ar$ = 1.0 to 0.1). The curves show the episodic return achieved by the DRL agents over 3000 training episodes for each $Ar$, reflecting the learning progress and control policy convergence across different geometrical configurations.}
    \label{fig:figure11}
\end{figure*}

We apply DRL-based flow control to elliptical cylinders with aspect ratios $Ar = 1$, $0.75$, $0.5$, $0.25$, and $0.1$. The corresponding learning curves are presented in \Cref{fig:figure11}, demonstrating that the effectiveness and stability of the DRL flow control strategy are significantly influenced by the geometry of the cylinders, with variations in $Ar$ playing a crucial role.

For the elliptical cylinders with $Ar = 1$ and $Ar = 0.75$, the DRL training learning curves converge around 1500 episodes. To demonstrate the stability and convergence performance of the training, we extended the training by an additional 1500 episodes. The step-like reward function curves in the later stages indicate that the RL agent effectively learned and stabilized the optimal control strategy.
In the case of the elliptical cylinder with $Ar = 0.5$, a rapid increase in the reward is initially observed, reaching a peak before gradually stabilizing. 
Compared to the cases with $Ar = 1$ and $Ar = 0.75$, the reward function exhibits similar convergence and stability.
As the $Ar$ decreases to 0.25 and 0.1, the learning curves of the DRL training slow down, and the stability deteriorates.
For elliptical cylinders with $Ar = 0.25$ and $Ar = 0.1$, the reward function exhibits noticeable fluctuations and delayed convergence, highlighting the difficulty of achieving a stable and effective control strategy at these extreme aspect ratios. 
Similar to the case with a $\beta$ of 0.24, the learning curves presented above indicate that the RL agent encounters greater challenges when performing flow control on more slender elliptical cylinders.

\subsubsection{Suppress fluid forces}

\begin{figure*}[hbt!]
    \centering
    \includegraphics[width=\textwidth]{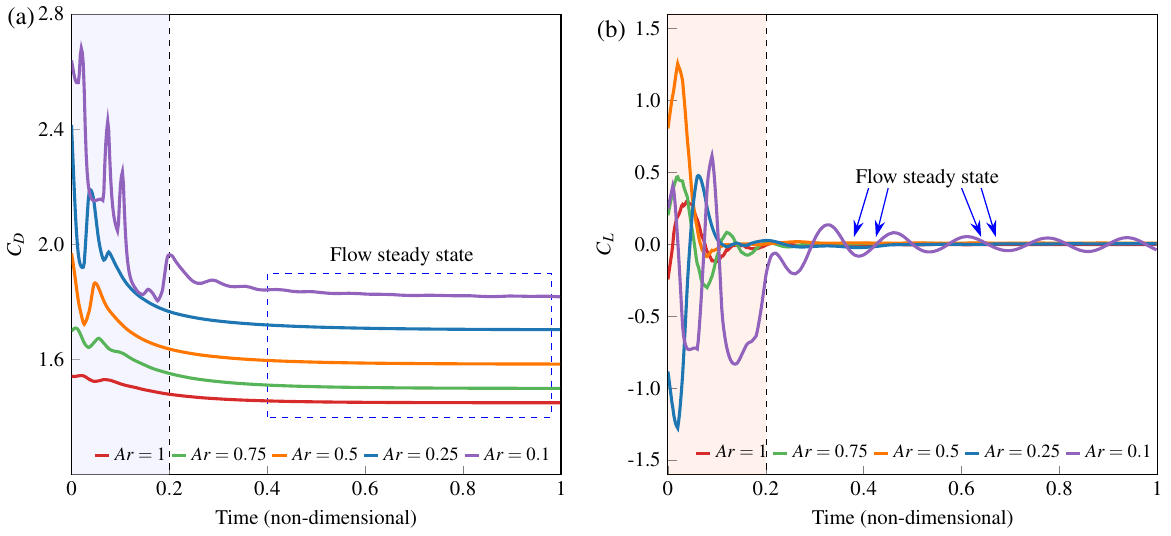}
    \caption{Dynamic response of the drag and lift coefficients for an elliptical cylinder with varying aspect ratios after the application of active flow control.
    (a) Time evolution of the drag coefficient for an elliptical cylinder subjected to AFC across a range of $Ar$ from 0.1 to 1. The plot illustrates how the drag coefficient decreases and approaches a steady state as time progresses for each $Ar$. The dashed vertical line indicates the time at which the flow reaches a steady state.
    (b) Time evolution of the lift coefficient for the same elliptical cylinder and $Ar$ as in the drag coefficient plot. This plot shows the initial fluctuations in lift coefficient followed by a stabilization towards a steady state, highlighted by the dashed vertical line.}
    \label{fig:figure12}
\end{figure*}

This section examines the drag reduction effect of flow control on elliptical cylinders with aspect ratios of $Ar = 1$, 0.75, 0.5, 0.25, and 0.1.
\Cref{fig:figure12} shows the time history curves of the drag coefficient and lift coefficient for these elliptical cylinders under controlled conditions.
For all five aspect ratios, the drag coefficient decreases significantly under flow control, quickly reaching its minimum value and then stabilizing. 
The drag coefficient of elliptical cylinders with all aspect ratios effectively reduces to a stable value, which differs from the value observed at $\beta = 0.24$. 
As the $\beta$ decreases, the influence of the sidewalls on the wake dynamics of the elliptical cylinder becomes less significant. This reduction in confinement allows even the slender elliptical cylinder with $Ar = 0.1$ to achieve a substantially reduced drag coefficient that remains nearly constant.

Similar to the drag coefficient control, the instantaneous response of the lift coefficient to flow control shows a significant reduction, rapidly reaching its minimum value after control is implemented. For elliptical cylinders with $Ar = 1$, 0.75, 0.5, and 0.25, the lift coefficient remains close to zero in the later stages of controlled flow, indicating complete suppression of lift, which is sustained over an extended period. However, for the elliptical cylinder with $Ar = 0.1$, the lift coefficient in the controlled flow proves more challenging to fully suppress to zero, highlighting the difficulty of achieving complete lift suppression in highly slender geometries.

\subsubsection{Flow structure comparison}\label{Flow structure02}

\begin{figure*}[htb!]
    \includegraphics{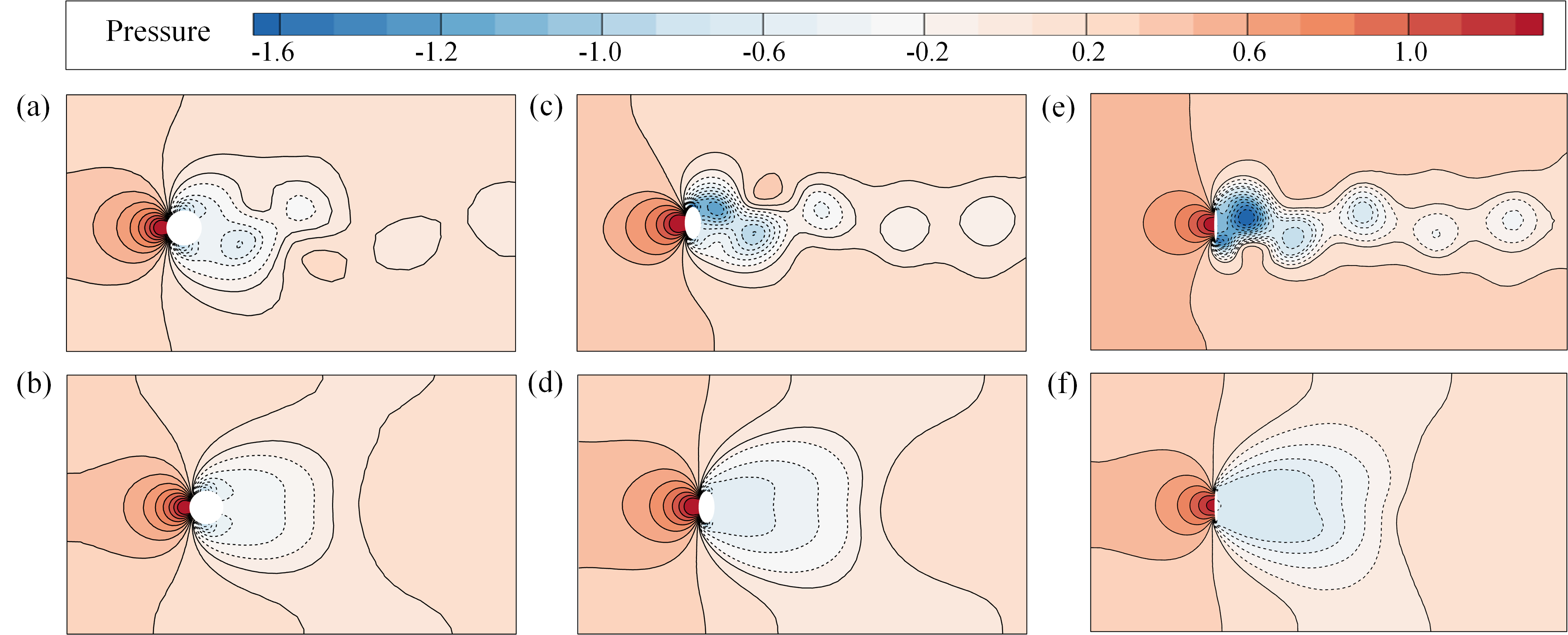}
    \caption{Instantaneous pressure contours before (top) and after (bottom) flow control for elliptical cylinders with $Ar$ = 1, 0.5, and 0.1.
    (a) Baseline flow for $Ar = 1$. (b) Controlled flow for $Ar = 1$. (c) Baseline flow for $Ar = 0.5$. (d) Controlled flow for $Ar = 0.5$. (e) Baseline flow for $Ar = 0.1$. (f) Controlled flow for $Ar = 0.1$.}
    \label{fig:figure13}
\end{figure*}

In \Cref{fig:figure13}, we show the instantaneous pressure contours for baseline and controlled flows of elliptical cylinders with $Ar = 1$, $0.5$, and $0.1$. 
For the cylinder at $Ar = 1$, localized irregular low-pressure zones are evident in the wake region. After the application of synthetic jet control, the pressure distribution in the wake becomes symmetrical and uniform, a pattern that is also noticeable for $Ar = 0.5$ and $0.1$. For the highly elongated cylinder with $Ar = 0.1$, a significant low-pressure region develops behind the cylinder in the baseline flow; however, in the controlled flow, this low-pressure zone transforms into a symmetrical shape, and the maximum negative pressure is significantly reduced.
This is clearly evident in the pressure contours, where the wake area behind the cylinder is significantly enlarged under active control compared to the baseline flow. This alteration results in a reduced average pressure drop within the wake, thereby contributing to the observed drag reduction.

Overall, the DRL-based flow control strategy demonstrates its effectiveness across elliptical cylinders with aspect ratios ranging from $Ar = 1$ to $Ar = 0.1$. This is evident in the significant suppression of the drag coefficient, with the periodic fluctuations in the drag coefficient being fully controlled to a stable state. 
Additionally, the pressure distribution around the elliptical cylinders transitions from an uneven state to a fully symmetric configuration. 
These results demonstrate the effectiveness of the proposed control approach for elliptical cylinders with aspect ratios ranging from $Ar = 1$ to $Ar = 0.1$.

\subsubsection{Eliminating vortices}

\begin{figure*}[hbt!]
    \includegraphics{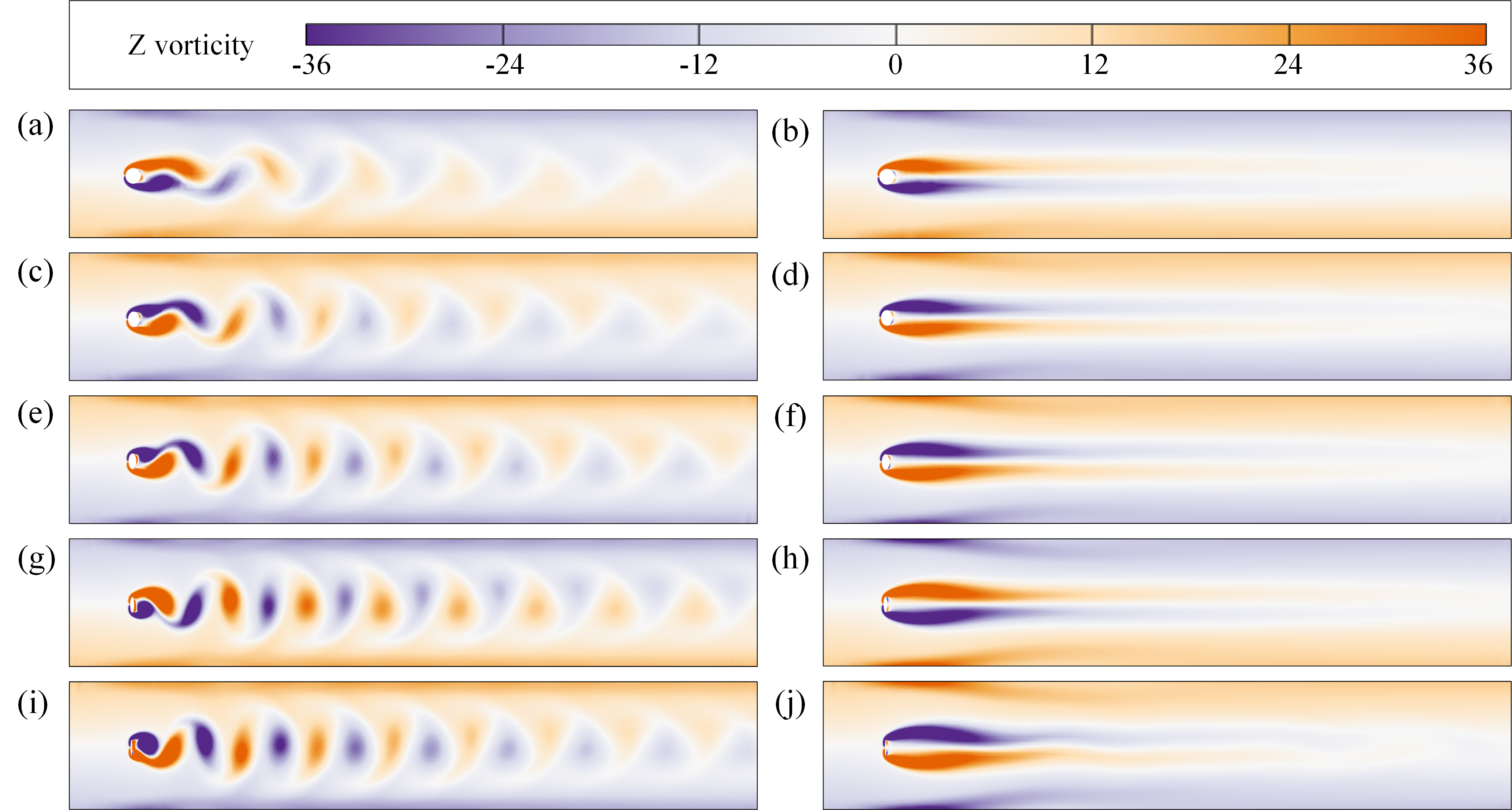}
    \caption{Comparison of instantaneous vorticity contours before and after flow control for elliptical cylinders with varying $Ar$. The left panels in each pair represent the uncontrolled baseline flow, while the right panels depict the controlled flow achieved via synthetic jet actuation. (a) Baseline flow for $Ar = 1.0$. (b) Controlled flow for $Ar = 1.0$. (c) Baseline flow for $Ar = 0.75$. (d) Controlled flow for $Ar = 0.75$. (e) Baseline flow for $Ar = 0.5$. (f) Controlled flow for $Ar = 0.5$. (g) Baseline flow for $Ar = 0.25$. (h) Controlled flow for $Ar = 0.25$. (i) Baseline flow for $Ar = 0.1$. (j) Controlled flow for $Ar = 0.1$.}
    \label{fig:figure14}
\end{figure*}

In this section, we investigate the efficacy of the DRL-based control strategy in suppressing vortex shedding for elliptical cylinders with varying $Ar$. 
\Cref{fig:figure14} presents the instantaneous vorticity contours around the elliptical cylinders, comparing the baseline flow conditions with those after the implementation of flow control.
In the baseline flow, all five elliptical cylinders ($Ar = 1$, 0.75, 0.5, 0.25, and 0.1) exhibit periodic vortex shedding, with the shedding frequency increasing as the aspect ratio decreases. 
After the flow control is applied, vortex shedding is completely suppressed in the controlled flow for all aspect ratios, demonstrating the effectiveness of the control strategy in stabilizing the wake dynamics. 

For $\beta = 0.24$, the flow control strategy derived from DRL fails to achieve complete suppression of vortex shedding in the wake of elliptical cylinders with $Ar = 0.5$, 0.25, and 0.1.
This suggests that the control strategy may face challenges in mitigating wake instabilities under higher blockage ratio conditions, particularly for more slender elliptical cylinders.
In contrast, when $\beta$ is reduced to 0.12, vortex shedding is completely suppressed across all aspect ratios from $Ar = 0.1$ to $Ar = 1$. This highlights the enhanced control effectiveness at lower blockage ratios, indicating that the control strategy is more capable of stabilizing the wake dynamics as the flow conditions become less constrained.

\begin{figure*}[htb!]
    \includegraphics{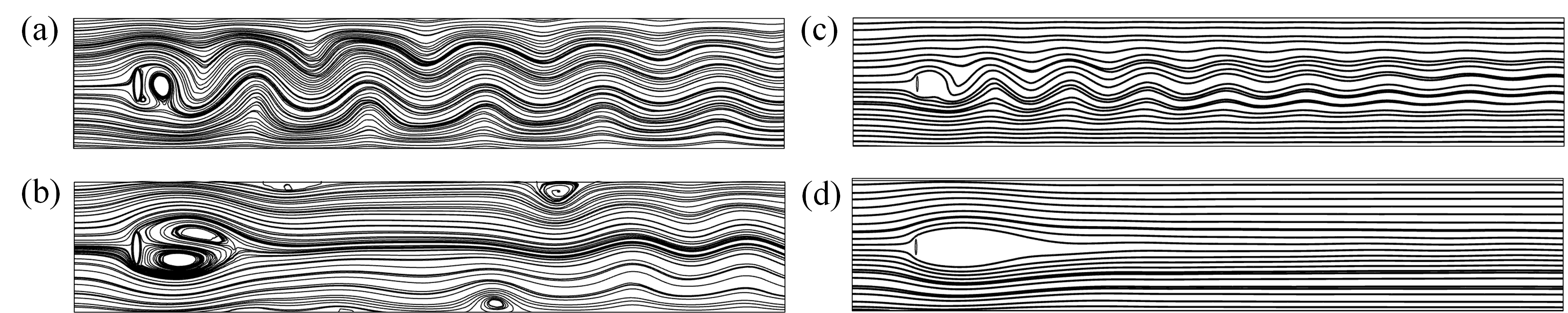}
    \caption{Comparison of the streamline distribution between the baseline flows and controlled flows for an elliptical cylinder with $ Ar = 0.1 $ at two different $\beta$. 
    (a) Baseline flow with $\beta = 0.24$. (b) Controlled flow with $\beta = 0.24$. (c) Baseline flow with $\beta = 0.12$. (d) Controlled flow with $\beta = 0.12$.}
    \label{fig:figure15}
\end{figure*}

We present the instantaneous streamline distribution for the elliptical cylinder with $Ar = 0.1$ under two blockage ratios in \Cref{fig:figure15}, comparing the baseline flow and the controlled flow.
By contrasting the instantaneous streamlines at blockage ratios of 0.24 and 0.12, we observe that in the baseline flow, the streamlines exhibit significant oscillations as the fluid flows past the cylinder.
When the $\beta$ is 0.24, the controlled flow still displays oscillatory behavior in the wake, with vortices forming near the upper and lower walls.
However, when the $\beta$ is reduced to 0.12, the controlled flow shows a stable wake structure. 
The streamlines remain parallel to the upper and lower boundaries, without exhibiting oscillations.
Notably, even for such a slender geometry, the DRL-based adaptive flow control successfully develops a strategy that completely suppresses vortex shedding.

\begin{figure*}[htb!]
    \centering
    \includegraphics{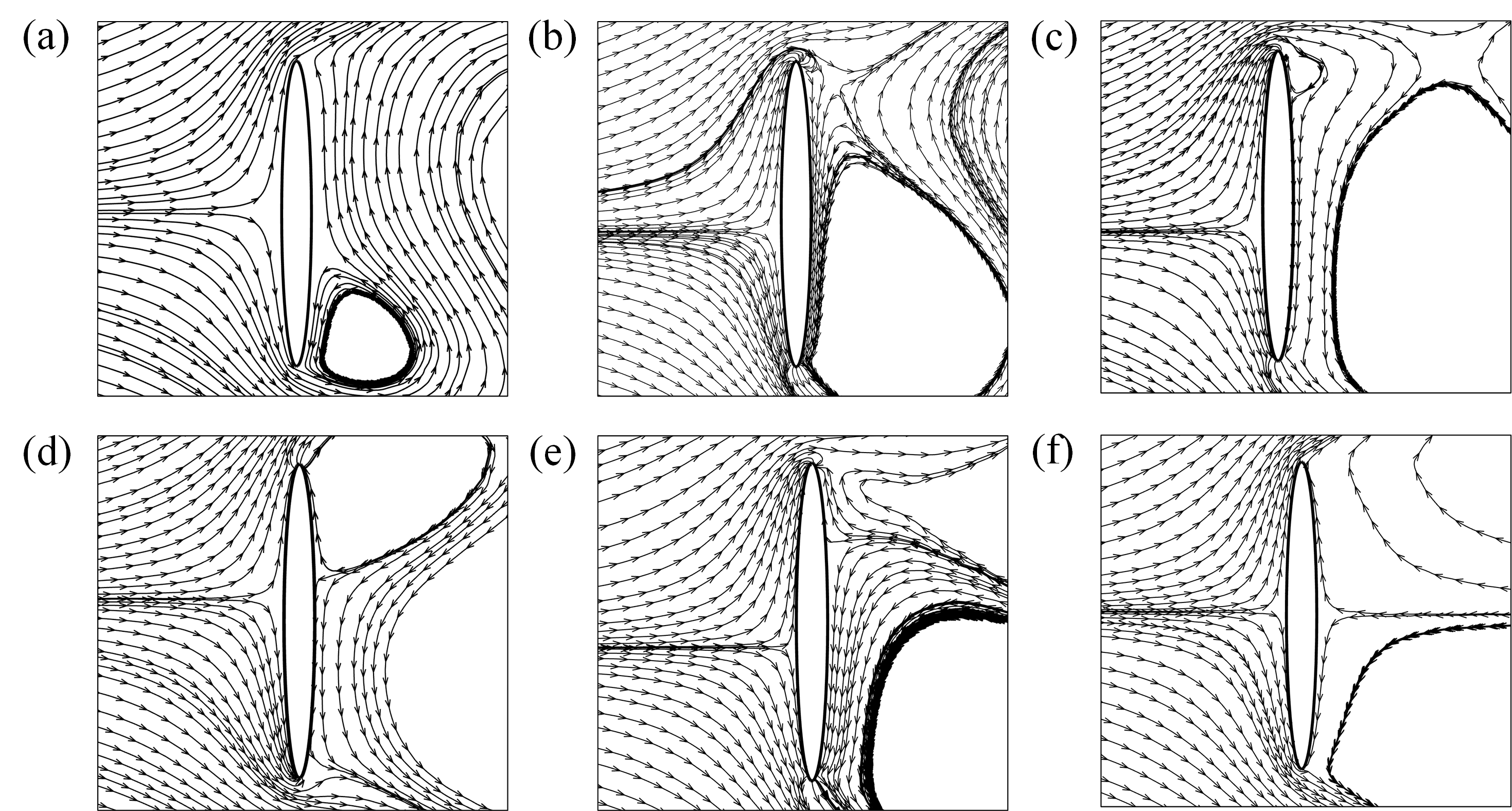}
    \caption{Time-resolved evolution of streamlines around an elliptical cylinder with aspect ratio $Ar = 0.1$ during the active flow control process. (a) Initial state at the onset of control ($t_0$), prior to actuation. (b) Activation phase of synthetic jet actuators at time $t_1$, initiating flow modification. (c–e) Intermediate control stages at times $t_2$, $t_3$, and $t_4$, illustrating the progressive suppression of vortex shedding and flow stabilization. (f) Final controlled state at time $t_5$, where a quasi-steady wake structure is established.}
    \label{fig:figure16}
\end{figure*}

\Cref{fig:figure16} shows the instantaneous streamlines around the elliptical cylinder with an aspect ratio of 0.1 and a $\beta$ of 0.12, captured before, during, and after flow control.
At time $t_0$, in the baseline flow, flow separation occurs at the upper and lower separation points of the elliptical cylinder, forming large separation bubbles. This initial state highlights the presence of significant flow separation before control is applied.
When flow control is initiated at time $t_1$, the synthetic jet at $90^\circ$ generates a suction force, while the jet at $270^\circ$ generates a blowing force. This combination causes the small vortex behind the cylinder to grow into a larger vortex. Subsequently, at time $t_2$, a new small vortex starts to emerge near the upper corner of the cylinder, indicating the dynamic response of the flow to the control forces.

When the flow control develops to time $t_3$, the synthetic jet at $90^\circ$ switches from suction to blowing, while the jet at $270^\circ$ switches from blowing to suction. This reversal leads to the development of the small vortex near the upper corner into a larger structure. As the vortices grow, they begin to compress against each other, resulting in intensified interactions between the upper and lower corner vortices.
By time $t_4$, the synthetic jet near $90^\circ$ resumes suction, while the jet at $270^\circ$ returns to blowing. Under the alternating influence of these synthetic jets, the vortices near the upper and lower corners gradually expand to their maximum size as they seek a state of equilibrium.
At time $t_5$, flow separation is observed near the $90^\circ$ and $270^\circ$ positions, leading to the formation of two symmetric vortices on the rear side of the cylinder. The streamlines around the cylinder indicate a stable and controlled wake structure.

\subsubsection{Energy consumption}

The DRL-based flow control strategy exhibits outstanding control performance, as demonstrated in the previous section. However, achieving energy efficiency while maintaining effective control remains a key objective. To evaluate the balance between control effectiveness and energy consumption, \Cref{tab:table3} presents the control performance and energy usage for elliptical cylinders with various $Ar$ at a $\beta$ of 0.12.
For the cylinder with $Ar = 1$, the DRL-based flow control strategy achieves a drag reduction of $6.1\%$. As the aspect ratio decreases, the drag reduction effect becomes more pronounced. Specifically, for the elliptical cylinder with $Ar = 0.75$, the reduction increases to $11.7\%$, while for $Ar = 0.5$, it further rises to $19.2\%$. For more slender cylinders, the drag reduction becomes even more significant. The control strategy achieves reductions of $27.8\%$ and $32.2\%$ for elliptical cylinders with $Ar = 0.25$ and $Ar = 0.1$, respectively.
The controlled flow effectively suppresses the lift coefficient for elliptical cylinders with $Ar$ ranging from 1 to 0.1, with the lift coefficient approaching zero in all cases. 
This indicates that the control strategy successfully mitigates lift coefficient fluctuations across the entire range of $Ar$ studied.
As the $Ar$ decreases, the suppression of lift remains effective, with the controlled lift coefficient consistently maintained at a relatively low level. 
This highlights the robustness of the control strategy in stabilizing both drag and lift coefficient.

\begin{table*}[hbt!]
\centering
\caption{Quantitative evaluation of control performance and energy consumption for elliptical cylinders with varying $Ar$ at blockage ratio $\beta = 0.12$. 
For each configuration at $Re = 100$, the baseline and controlled time-averaged drag coefficients ($\overline{C}_{D}$), drag reduction rate ($R_D$), and controlled lift coefficient ($\overline{C}_{L}$) are reported. The time-averaged actuation amplitude ($\overline{a}$) and its normalized percentage relative to the baseline are included to assess the energetic cost of flow control.
}
  \begin{tabularx}{\textwidth}{
    >{\centering\arraybackslash}p{0.06\linewidth}
    >{\centering\arraybackslash}p{0.06\linewidth}
    >{\centering\arraybackslash}p{0.11\linewidth}
    >{\centering\arraybackslash}p{0.11\linewidth}
    >{\centering\arraybackslash}p{0.11\linewidth}
    >{\centering\arraybackslash}p{0.12\linewidth}
    >{\centering\arraybackslash}p{0.08\linewidth}  
    >{\centering\arraybackslash}p{0.13\linewidth}     
  }
\toprule 
    $Ar$ & $Re$ & $\overline{C}_{D,\text{Baseline}}$ & $\overline{C}_{D,\text{Controlled}}$ & $R_D$ (\%) &  $\overline{C}_{L,\text{Controlled}}$ & $\overline{a}$ & $\overline{a}$ ratio (\%) \\
    \midrule
    1.00    & 100 & 1.543 & 1.449 & 6.1   & -0.002 & -0.013 & 1.3  \\
    0.75    & 100 & 1.699 & 1.500 & 11.7  & 0.005  & -0.025 & 2.5 \\
    0.50    & 100 & 1.960 & 1.584 & 19.2  & 0.008  & -0.021 & 2.1 \\
    0.25    & 100 & 2.360 & 1.704 & 27.8  & 0.003  & 0.028 & 2.8 \\
    0.10    & 100 & 2.685 & 1.820 & 32.2  & -0.003 & 0.162 & 16.2  \\
\bottomrule
\end{tabularx}
\label{tab:table3}
\end{table*}

We use the average synthetic jet mass flow rate to represent the external energy required to maintain flow control performance in the later stages of the process. The $\overline{a}$ ratio (\%) is defined as the ratio of the synthetic jet mass flow rate to the upstream inlet flow rate, serving as a measure of the energy consumption.
For the elliptical cylinder with $Ar = 1$, maintaining stable control requires only 1.3\% of the upstream inlet flow rate as external energy. This indicates a highly energy-efficient control strategy.
For cylinders with aspect ratios ranging from $Ar = 0.75$ to $Ar = 0.25$, the external energy required to sustain control performance remains within 3\% of the inlet flow rate. This further demonstrates the energy efficiency of the control approach.
However, for the most slender cylinder, the active control demands significantly more energy. The ratio of the synthetic jet mass flow rate to the inlet flow rate reaches 16.1\%, highlighting the increased energy requirements associated with slender geometries.

\begin{figure*}[htb!]
    \centering
    \includegraphics[width=\textwidth]{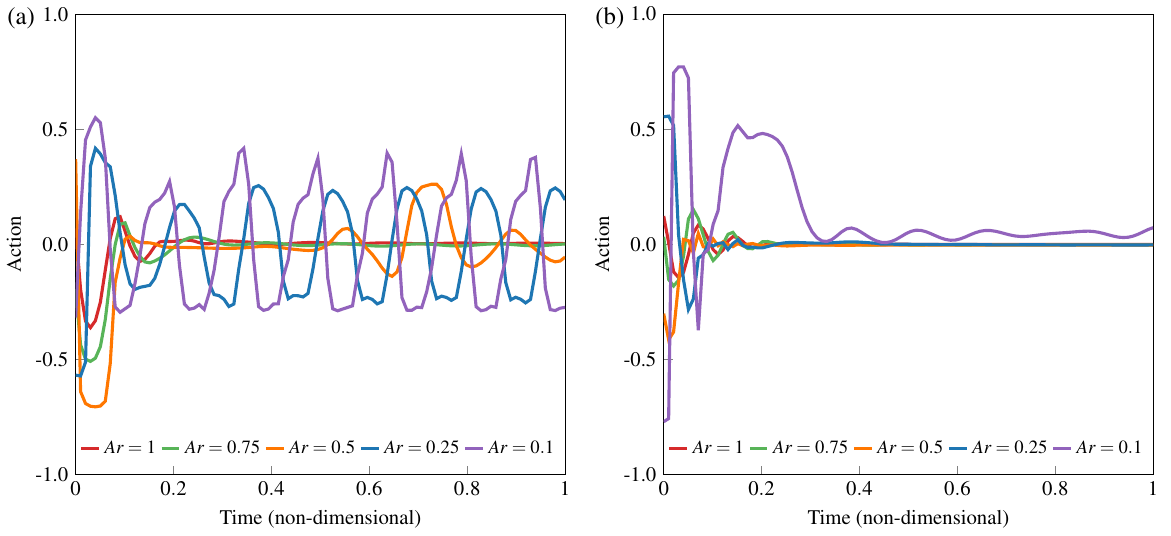}
    \caption{Energy consumption associated with flow control of an elliptical cylinder across varying aspect ratios ($Ar = 1.0$ to $0.1$) under different blockage ratios. 
    (a) Case of moderate confinement with $\beta = 0.24$, illustrating the energy input required for synthetic jet actuation at each $Ar$. 
    (b) Case of lower confinement with $\beta = 0.12$, highlighting the reduced energy demand due to weaker wall-induced interactions. 
    These results reflect the trade-off between control performance and energetic cost across different geometric and confinement configurations.}
    \label{fig:figure17}
\end{figure*}

To facilitate the comparative analysis of flow control strategies for elliptical cylinders with different aspect ratios under $\beta$ of $ 0.24 $ and $ 0.12 $, \Cref{fig:figure17} illustrates the temporal evolution of the jet actuation values. 
This comparison helps to understand how geometric parameters, particularly the $ Ar $ and $\beta$, influence the effectiveness of DRL-based flow control strategies.
The time series of synthetic jet control actions provides insight into the control strategy’s performance, where lower action amplitudes indicate higher cost efficiency and minimal fluctuations in action values reflect effective flow control. 
The evolution of control actions for elliptical cylinders with various $Ar$ during the control period, when $\beta$ is 0.24, is shown in \Cref{fig:figure17} (a).
For elliptical cylinders with $Ar = 1$ and $Ar = 0.75$, the flow control starts with a high jet velocity. This velocity then decreases to a stable minimum. The initial high jet velocity effectively suppresses the flow, while the lower velocity maintains stability.
As $Ar$ decreases and the cylinders become more slender, the control strategies change significantly. The fluctuations in control actions increase, reflecting the challenges posed by vortex shedding.

\Cref{fig:figure17} (b) illustrates the evolution of control actions during flow control for elliptical cylinders of varying $Ar$ at $\beta = 0.12$.
The control strategies for elliptical cylinders with $Ar$ ranging from $Ar = 1$ to $Ar = 0.25$ exhibit a consistent pattern.
In the initial phase of flow control, the control actions are characterized by large amplitude fluctuations, reflecting substantial energy input. 
As the control process advances, the amplitude of these actions decreases rapidly, indicating a transition to a more energy-efficient state. 
In the later stages, the control actions reach a steady state, ensuring sustained flow control.
Notably, the energy expenditure associated with control increases progressively as the aspect ratio decreases, highlighting the heightened demand for energy in stabilizing slender geometries.
This control strategy is consistent for elliptical cylinders with $Ar$ from $Ar = 1$ to $Ar = 0.25$ under $\beta = 0.12$.
Similar control patterns have also been reported in studies by Rabault et al. \cite{rabault2019artificial} and Li et al. \cite{liReinforcementlearning}.
Although this is more challenging for extremely slender cylinders with $Ar = 0.1$, the control strategy still maintains relatively low energy consumption.
Compared to the scenario under $\beta = 0.24$, the control strategy at $\beta = 0.12$ demonstrates significantly higher energy efficiency.

\section{Conclusions}\label{sec:Conclusions}

This study applied an advanced DRL algorithm to develop an active flow control strategy for complex flow systems.
A DRL framework was constructed to adjust the flow rates of synthetic jets mounted on an elliptical cylinder in real time, effectively regulating the wake structure.
Specifically, the study investigated the control performance of the DRL-based strategy for flows around elliptical cylinders with aspect ratios of $Ar = 1.0$, $0.75$, $0.5$, $0.25$, and $0.1$, under blockage ratios of $\beta = 0.24$ and $0.12$, respectively.

Our results demonstrated that the effectiveness of the DRL-based flow control strategy was highly sensitive to both the $Ar$ and the $\beta$ of the elliptical cylinder.
With a relatively high blockage ratio of $\beta = 0.24$, effective suppression of vortex shedding was achieved for configurations with $Ar = 1.0$, $0.75$, and $0.5$, with the control input requiring only about 1\% of the inlet flow energy.
In contrast, for more slender bodies with $Ar = 0.25$ and $0.1$, the DRL control algorithm managed to achieve only a modest level of drag reduction. The effectiveness in suppressing vortex shedding remained limited, with the actuation energy cost rising sharply to 14.5\% of the inlet energy.
When the blockage ratio $\beta$ was reduced to $\beta = 0.12$, the DRL framework showed improved adaptability and robustness across all tested $Ar$ values. Even for the most slender cylinder ($Ar = 0.1$), the reward function converged and effective control strategies emerged.
Across all configurations, the learned policies led to substantial drag reduction, ranging from 6.1\% to 32.3\%, while maintaining near-zero lift fluctuations.
These findings highlighted a trade-off between geometry and control efficiency: as the aspect ratio decreased, achieving stable and energy-efficient control became increasingly challenging.

Nonetheless, our study demonstrated that DRL had strong potential in adapting to varied flow conditions, offering a data-driven solution for real-time flow control over a wide range of bluff body geometries at different aspect and blockage ratios.
Future research could integrate real-time sensor feedback and extend the control framework to more complex and unsteady flow environments, advancing the practical application of DRL in fluid mechanics.










\appendix
\setcounter{figure}{0}
\setcounter{table}{0}

\section{Details of simulation environment}\label{appA}
To facilitate the replication of the results from this study, a detailed explanation of the formal definitions of the boundary conditions and physical parameters involved in this research is provided.
At the inlet $\Gamma_\text{in}$, the inflow velocity along $x$-axis is prescribed by a parabolic velocity profile in the form,
\begin{equation}\label{equ:1velocity}
U_{\text{inlet}}(y) = U_m \frac{(H - 2y)(H + 2y)}{H^2}, \tag{A1}
\end{equation}
where $ U_m $ is the maximum velocity magnitude of the parabolic profile, which is set to 1.5 in this study, and $ H=4.1D $ represents the total height of the rectangular domain. Additionally, the velocity along the y-axis, $ V_{\text{inlet}}(y) = 0 $. The average inlet velocity $\overline{U}$, is related to the parabolic velocity profile $U_{\text{inlet}}(y)$ through the expression:
\begin{equation}\label{equ:mean velocity}
\overline{U}=\frac{1}{H} \int_{-H/2}^{H/2} U_{\text{inlet}}(y) d y=\frac{2}{3} U_m=1. \tag{A2}
\end{equation}
We define $ f_{Q_i} = A(\theta; Q_i)(x, y) $, where the modulation depends on the angular coordinate $ \theta $ as shown in \Cref{fig:figure1}(b). For the jet, which has a width $ \omega $ and is centered at $ \theta_0 $ on the cylinder with radius $ R $, the modulation is given by:
\begin{equation}
A(\theta; Q) = Q \frac{\pi}{2 \omega R^2} \cos\left(\frac{\pi}{\omega} (\theta - \theta_0)\right), \tag{A3}
\end{equation}
where $ Q $ represents the amplitude of the modulation, $ \omega $ denotes the width of the jet, $R$ is the radius of the cylinder, $ \theta_0 $ is the angular position of the jet center, and $ \theta $ is the angular coordinate along the cylinder surface. The cosine function governs the spatial variation of the modulation along the angular coordinate. This modulation function smoothly integrates with the no-slip boundary conditions on the cylinder surface.

\setcounter{figure}{0}
\setcounter{table}{0}
\section{Grid independence test}\label{appB}

\begin{figure}[htb!]
    \centering
    \includegraphics{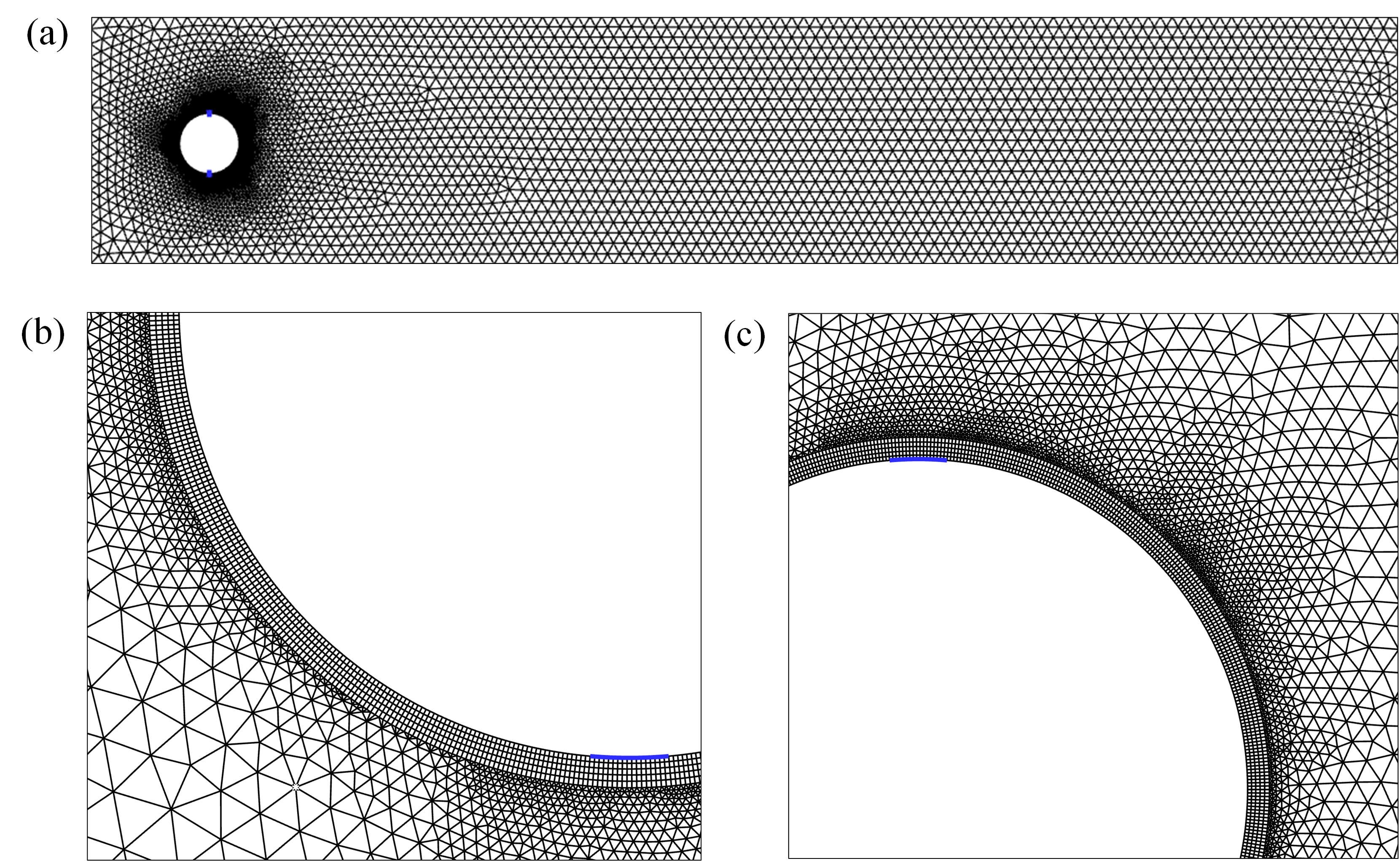}
    \caption{Numerical mesh refinement and synthetic jet actuator placement for elliptical cylinder flow control. (a) Overview of the entire computational domain, highlighting the elliptical cylinder and the positioning of synthetic jet actuators. (b) Detailed view of the mesh near the cylinder, showing localized refinement to resolve flow–actuator interactions. (c) Close-up of the immediate vicinity of the cylinder, illustrating the actuator configuration and fine mesh resolution essential for accurate flow control simulations.}
    \label{fig:figureB1}
\end{figure}

\Cref{fig:figureB1} illustrates the discretization of the computational domain for the elliptical cylinder with an aspect ratio of $Ar = 1$ in the medium-scale configuration.
\Cref{tab:tableB1} presents the simulation results obtained using three different mesh resolutions, comparing the maximum drag coefficient $C_{D,\text{max}}$, mean drag coefficient $C_{D,\text{mean}}$, maximum lift coefficient $C_{L,\text{max}}$, and Strouhal number $St$.
To ensure the validity of the numerical method, the simulation results are compared against benchmark values reported by Schäfer et al. \cite{schafer1996benchmark}. Additionally, we compared our calculation results with pioneering research results \cite{rabault2019artificial}.

\begin{table}[htb!]
\centering
\caption{Mesh convergence study for the two-dimensional flow past an elliptical cylinder with $Ar = 1$ at $Re = 100$. The results show consistency with previous studies and confirm the reliability of the present simulations through mesh-independent flow parameters.}
\begin{tabular}{
  >{\centering\arraybackslash}p{0.18\textwidth}
  >{\centering\arraybackslash}p{0.14\textwidth}
  >{\centering\arraybackslash}p{0.08\textwidth}
  >{\centering\arraybackslash}p{0.11\textwidth}
  >{\centering\arraybackslash}p{0.1\textwidth}
  >{\centering\arraybackslash}p{0.11\textwidth}
  >{\centering\arraybackslash}p{0.11\textwidth} 
}
\toprule
Configuration & Mesh resolution & Mesh & $C_{D, max}$ & $C_{D, mean}$ & $C_{L, max}$ & $St$ \\
\midrule
Schafer et al. \cite{schafer1996benchmark} & - & - & 3.220–3.240 & - & 0.990–1.010 & 0.295–0.305 \\
Rabault et al. \cite{rabault2019artificial} & - & 9262 &  & 3.205 & - &  \\
          & {Coarse}   & 9,460  & 3.248 & 3.223 & 1.016 & 0.305 \\
$Ar=1$    & {Medium}   & 16,452 & 3.226 & 3.207 & 1.032 & 0.302 \\
          & {Fine}     & 22,348 & 3.229 & 3.208 & 1.033 & 0.302 \\
\bottomrule
\end{tabular}
\label{tab:tableB1}
\end{table}

The simulation results obtained with the coarse mesh exhibit significant discrepancies in drag and lift coefficients compared to the results of Rabault et al. \cite{rabault2019artificial}.
In contrast, the medium and fine mesh resolutions yield results that fall within the upper and lower limits of the standard benchmark \cite{schafer1996benchmark}, with the computational errors relative to Rabault et al.\cite{rabault2019artificial} remaining below 0.1\%.
This indicates that both the medium and fine meshes are capable of accurately capturing the flow characteristics around the elliptical cylinder.
However, the fine mesh, despite its superior resolution, incurs a substantial computational cost due to the increased number of grid points, significantly impeding the training speed of reinforcement learning. This trade-off between accuracy and computational efficiency necessitates a careful selection of the mesh resolution \cite{rabaultAccelerating,jia2024optimal,Font2025turbulent}.
Given that the medium mesh offers comparable accuracy to the fine mesh while substantially reducing computational demands, it represents a more balanced choice for subsequent numerical simulations and reinforcement learning training.
Therefore, we adopt the medium-scale mesh scheme to ensure both accuracy and efficiency.

\setcounter{figure}{0}
\setcounter{table}{0}
\section{Proximal Policy Optimization algorithm}\label{appC}

Proximal Policy Optimization algorithm is a policy gradient method in deep reinforcement learning that optimizes an agent's policy by performing gradient ascent on the expected cumulative reward. It adjusts policy parameters to map states to actions, maximizing rewards effectively. 

\begin{itemize}

\item \textbf{PPO objective function.}
The core objective of PPO is formulated as:
\begin{equation}
L^{\text{PPO}}(\theta) = \mathbb{E}_t \left[ \min \left( r_t(\theta) \hat{A}_t, \text{clip}\left(r_t(\theta), 1 - \epsilon, 1 + \epsilon\right) \hat{A}_t \right) \right],  \tag{C1}
\end{equation}
where $ r_t(\theta) = \frac{\pi_\theta(a_t|s_t)}{\pi_{\theta_{\text{old}}}(a_t|s_t)} $ represents the probability ratio between the new policy $ \pi_\theta $ and the old policy $ \pi_{\theta_{\text{old}}} $. 
The advantage function $\hat{A}_t$ guides policy updates, while the clipping parameter $\epsilon$ prevents large updates. The objective balances $ r_t(\theta) \hat{A}_t $ for improvement and $\text{clip}(r_t(\theta), 1-\epsilon, 1+\epsilon)$ for stability.

\item \textbf{Parameter update.}
PPO iteratively updates the policy parameters $ \theta $ using gradient ascent, following the update rule 
\begin{equation}
\theta_{k+1} = \theta_k + \alpha \nabla_\theta L^{\text{PPO}}(\theta_k), \tag{C2}
\end{equation}
where $ \theta_{k+1} $ represents the updated policy parameters after the $ k $-th iteration, and $ \theta_k $ denotes the parameters before the update. The learning rate $ \alpha $ determines the step size for each update, while $ \nabla_\theta L^{\text{PPO}}(\theta_k) $ is the gradient of the PPO objective function with respect to the policy parameters at the $ k $-th iteration.

\item \textbf{Iterative process.}
The algorithm operates through iterations that involve collecting trajectories using the current policy, estimating the advantages $\hat{A}_t$ to evaluate actions, and updating the policy by applying the gradient ascent rule. This process repeats until the policy converges, effectively balancing exploration and stability through PPO for optimal policy optimization. 

\end{itemize}

\setcounter{figure}{0}
\setcounter{table}{0}
\section{Hyperparameters}\label{appD}

The primary parameters used in the numerical simulations are detailed in \Cref{sec:Simulation}, followed by a comprehensive discussion of the hyperparameters of the DRL learning algorithm in \Cref{sec:DRL_AFC}.
These parameters and their specific settings are summarized in \Cref{tab:table5}.

\begin{table}[hbt!]
\centering
\caption{Key parameters and settings used in the CFD simulations and DRL--based control framework.}
\begin{tabular*}{\textwidth}{@{\extracolsep{\fill}} l l l }
\toprule
\textbf{Parameter} & \textbf{Symbol} & \textbf{Value} \\
\midrule
\multicolumn{3}{c}{\textbf{\textit{~~~~~Flow simulation set-up}}} \\
Reynolds number & $Re$ & 100  \\ 
Jet width (°) & $w$ & 10  \\ 
Numerical time step (non-dimensional) & dt & $5 \times 10^{-3}$  \\ 
The average inlet velocity (m/s) & U & 1  \\ 
The maximum inlet velocity (m/s) & $U_{\text{max}}$ & 1.5  \\ 
\midrule
\multicolumn{3}{c}{\textbf{\textit{~~~~~DRL Hyperparameters}}} \\
Episode number & - & 3000  \\
Learning rate & $lr$ & 0.001  \\ 
Discount factor & $\gamma$ & 0.99  \\ 
Policy network & $\pi_\theta$ & 512*512  \\ 
Policy Ratio Clipping & $\epsilon$ & 0.2  \\ 
Optimizer  & - & Adam \\ 
Batch size & - & 64  \\ 
Steps per episode & Timesteps & 100  \\ 
Interaction period (non-dimensional) & $T_{max}$ & 2.5  \\ 
\bottomrule
\end{tabular*}
\label{tab:table5}
\end{table}

\section*{Data availability}
Data will be made available on request.

\bibliographystyle{elsarticle-num}
\bibliography{elsarticle}

\end{document}